\documentclass{jaa}
\usepackage{natbib}
\usepackage[dvipsnames]{xcolor}
\usepackage{graphicx}
\usepackage{newtxtext,newtxmath}

\usepackage[T1]{fontenc}
\usepackage{csquotes}
\usepackage{amsmath}	
\usepackage{amssymb}	
\usepackage{txfonts}
\usepackage{textcase}
\usepackage{balance}
\usepackage{float}
\usepackage{xspace}
\usepackage[colorlinks,citecolor=blue,urlcolor=blue,bookmarks=false,hypertexnames=true]{hyperref} 
\usepackage{array, booktabs, makecell, multirow}
\newcommand{\astro}{\textit{AstroSat}}
\newcommand{\src}{2S 1417--624}
\newcommand{\erg}{ergs cm$^{-2}$ s$^{-1}$}
\newcommand{\mnras}{Monthly Notices of the Royal Astronomical Society}
\newcommand{\apj}{The Astrophysical Journal}
\newcommand{\apjl}{The Astrophysical Journal}

\newcommand{\aap}{Astronomy and Astrophysics}

\newcommand{\apjs}{The Astrophysical Journal Supplement}
\newcommand{\pasj}{Publications of the Astronomical Society of Japan}
\newcommand{\ssr}{Space Science Reviews}
\newcommand{\pasp}{The Publications of the Astronomical Society of the Pacific}
\newcommand{\aaps}{Astronomy and Astrophysics Supplement}
\newcommand{\apss}{Astrophysics and Space Science}
\newcommand{\aapr}{The Astronomy and Astrophysics Review}
\newcommand{\aj}{The Astronomical Journal}

\begin{document}\sloppy

\title{The 2021 outburst of \src\ revisited with \astro}

\author{Chetana Jain\textsuperscript{1,*}}
\affilOne{\textsuperscript{1}Hansraj College, University of Delhi, Delhi 110007, India.\\}

\twocolumn[{
\maketitle
\corres{chetanajain11@gmail.com}

\begin{abstract} 
This work presents the first ever broadband (0.7--25.0 keV) timing and spectral analysis of Be-HMXB \src\ during its 2021 outburst. Using \astro\ observations, coherent pulsations at $\sim$17.36633 s (MJD 59239.082) were detected in 0.7-7.0 keV SXT and 3.0-25.0 keV LAXPC data. The pulse profile was dual peaked at all energies, with the relative intensity of main peak increasing with energy. The peaks in the SXT profiles were broad and comprised of several mini-structures. The LAXPC profiles were relatively smooth and had higher pulsed fraction which increased with energy. The SXT+LAXPC simultaneous energy spectrum is well described by an absorbed power-law with exponential cut-off, along with $\sim$1.6 keV black body component and 6.47 keV emission line. A model comprising of an absorbed power law with high energy cutoff plus a partial covering absorber and Gaussian emission line also fits the spectrum quite well. These results have been compared with timing and spectral features during the previous outbursts of this transient pulsar. 
\end{abstract}
\keywords{stars: neutron; X-rays: binaries; accretion; accretion disks; pulsars: individual: \src}
}]

\doinum{12.3456/s78910-011-012-3}
\artcitid{\#\#\#\#}
\volnum{000}
\year{0000}
\pgrange{1--}
\setcounter{page}{1}
\lp{14}

\section{Introduction}

Be/X-ray binaries (BeXBs) are a sub-class of high mass X-ray binaries (HMXBs) where the optical companion can be a dwarf, sub-giant or giant OBe star of luminosity class III-V and they exhibit line emission over the continuum spectrum \citep{Porter03, Reig11}. These systems are associated with radially extended equatorial disc, formed due to material expelled from rapidly rotating Be star. Recurrent outbursts occur due to episodic accretion of matter by the neutron star as it passes close to this disc \citep{Negueruela98, Rivinius13}. Based on factors such as the orientation of the disc with respect to the neutron star and the mass accretion rate, these outbursts vary in intensity and duration; thereby providing a wide range of observational probes to study these interesting objects.

\src\ is one such Be-HMXB which has undergone several X-ray outbursts since its discovery in 1978 with Small Astronomy Satellite (SAS-3) \citep{Apparao80}. During this first detection, the source exhibited hard spectrum and $\sim$17.64 s X-ray pulsations were detected \citep{Kelley81}. The 2.0-11.0 keV pulse profile comprised of sharp features along with an inter-pulse structure. Using the SAS-3 observations spanning about eight days, \citet{Kelley81} found that the neutron star was spinning up at a rate of (3–6)$\times$10$^{-11}$ Hz s$^{-1}$, possibly due to large intrinsic spin-up of the pulsar and its binary orbital motion about an early type giant star. Their results indicated a mass function $\ge$5M$_\odot$ and an orbital period $\ge$37 days. Later, based on optical observations from the \textit{Cerro-Tololo Inter-American Observatory} and X-ray observations with the \textit{Einstein X-ray Observatory}, \citet{Grindlay84} identified the optical counterpart to a Be star in a wide orbit just like other HMXBs such as 4U 0115+63. 

Since its discovery, occurrence of several outbursts in \src\ has been reported with various instruments on board different X-ray missions (1994: \citet{Finger96}; 1999: \citet{Inam04}; 2009: \citet{Beklen09}, \citet{Gupta18}; 2018: \citet{Gupta19}, \citet{Ji20} and 2021: \citet{Hazra21}). From the CGRO-BATSE observations of 1994 outburst, spanning almost 10 months, \citet{Finger96} reported the orbital parameters of \src\ which were later refined by \citet{Inam04} and \citet{Raichur10} by using the \textit{RXTE} observations of 1999 outburst. \citet{Raichur10} also traced the evolution of spin period as the outburst progressed and based on a considerable decrease in the estimated value of longitude of periastron, they provided hints towards a retrograde orbit.

During each outburst in \src, the neutron star is known to spin-up at a rate correlated with the X-ray flux \citep{Kelley81, Finger96, Inam04, Gupta18, Gupta19}. Such correlation in different energy ranges is known in several Be-HMXBs such as, EXO 2030+375 \citep{Parmar89}, A 0535+26 \citep{Bildsten97} and XTE J1543-568 \citep{Zand01}. From \textit{RXTE} observations of 1999 outburst, \citet{Inam04} found that even the morphology of the pulse profile and the pulsed fraction was correlated with the X-ray flux. This observation was validated during 2009 and 2018 outburst \citep{Gupta18, Gupta19}.
\citet{Inam04} observed that the pulse shape during increasing and decreasing flux and near the peak of the outburst were similar, with a slight variation in the pulsed fraction. However, the shape and pulsed fraction was considerably different at low flux values. \citet{Gupta18} and \citet{Gupta19} found that the pulse structure changed from double peak at low luminosity to triple peak at higher luminosity. The pulsed fraction showed anti-correlation with the source flux. These observations are in contrast to other BeXBs such as SAX J2103.5+4545 where both pulsed fraction and the pulse shape are independent of the X-ray flux level \citep{Baykal02}.

The first detailed spectral analysis of \src\ was reported from its 1999 outburst, during which the 3.0-55.0 keV \textit{RXTE}/PCA-HEXTE spectrum was described with a power law along with low energy absorption, an exponential high energy cut-off and a Gaussian component for 6.4--6.8 keV iron line complex \citep{Inam04}. The presence of complex iron line feature was attributed to cold iron atoms and H/ He like lines due to hot ionized gas around the pulsar. The 3.0-30.0 keV \textit{RXTE} spectrum during 2009 outburst also comprised of similar spectral components \citep{Gupta18}. An additional black body component was required to describe the combined \textit{NuSTAR} and \textit{Swift}-XRT 0.9-79.0 keV spectrum during 2018 outburst \citep{Gupta19}. Using \textit{NICER} data, \citet{Serim22} modeled the 0.2-12.0 keV spectrum with partial covering fraction absorption component instead of black body component. 

Recently, \citet{Mandal22} reported the only known analysis of the 2021 outbursts of \src\ using \textit{NICER} observation. The pulsar was spinning up at (0.8-1.8)$\times$10$^{-11}$ Hz s$^{-1}$ during this outburst. The pulse profile comprised of multiple peaks and dips which evolved with energy. During this observation, \src\ went through a state transition from the sub-critical to super-critical accretion regime. The energy spectrum was well described by a model comprising of a power-law with a higher cut-off energy and black body components along with a photo-electric absorption component. A 6.4 keV iron emission line with an equivalent width of 50 eV was also detected. 

In this paper, timing and spectral analysis of archival data of \src\ with \astro\ taken during 2021 outburst has been presented. Since this observation was taken during the rising phase of the outburst, the primary aim of this work is to study the broadband luminosity dependence of timing and spectral features and compare the results with those reported in previous outbursts. Section~\ref{sec:obs} gives the observation and data reduction details of \src, followed by timing and spectral analysis in Section~\ref{sec:analysis}. The results have been discussed in Section~\ref{sec:discuss}.

\section{Observation and Data Reduction}
\label{sec:obs}

The current analysis is based on data taken with \astro\ which was launched in September 2015 by the Indian Space Research Organisation \citep{Agrawal06, Singh14}. This satellite comprises of five payloads which can simultaneously observe in a wide band of X-rays, visible light and far ultraviolet radiations. These include Soft X-ray Telescope (SXT: 0.3--8.0 keV \citep{Singh17}), Large Area X-ray Proportional Counter (LAXPC: 3.0--80.0 keV \citep{Yadav16, Agrawal17}), Cadmium-Zinc-Telluride Imager (CZTI: 10.0--100.0 keV \citep{Rao17}), Ultra-Violet Imaging Telescope (UVIT: 130--550 nm \citep{Tandon17}) and Scanning Sky Monitor (SSM: 2.5--10.0 keV \citep{Ramadevi18}).

\begin{figure}
    \centering
    \includegraphics[width=0.7\columnwidth,angle=-90]{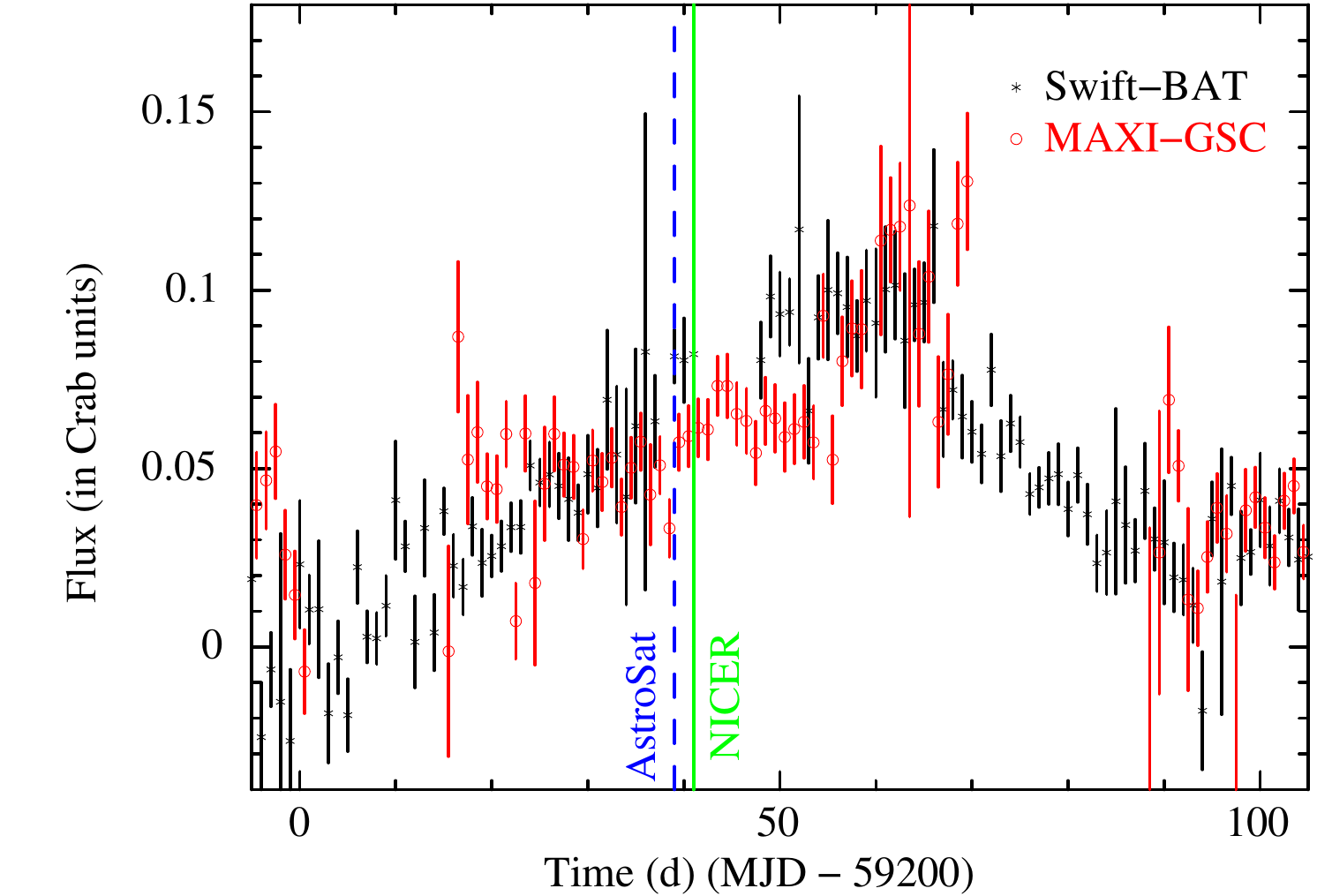}
    \caption{The 2--20 keV \textit{MAXI}-GSC (shown with red circled marker) and 15--50 keV \textit{Swift}-BAT (shown with black asterisk) light curve of \src. The dashed blue vertical line at MJD 59239 corresponds to the \astro\ observation analysed in this work. The green vertical line shows the epoch of \textit{NICER} observation used by \citet{Mandal22}.}
    \label{fig:longtermlc}
\end{figure}

\src\ entered the outburst phase around 2020 December 17 (MJD 59200) and lasted for about 100 days. Figure~\ref{fig:longtermlc} shows its long term 2--20 keV \textit{MAXI}-GSC light curve\footnote{\url{http://maxi.riken.jp/star_data/J1421-626/J1421-626.html}} \citep{Mihara11} overlaid on the 15-50 keV \textit{Swift}-BAT light curve\footnote{\url{https://swift.gsfc.nasa.gov/results/transients/weak/H1417-624/}} \citep{Barthelmy05, Krimm13}. The count rates have been normalised to respective Crab units \footnote{\url{http://maxi.riken.jp/top/readme.html , https://swift.gsfc.nasa.gov/results/transients/}}. The \textit{MAXI} count rate has been scaled up by a factor of 3 for better representation. The slight mismatch in these light curves at few places could be due to different energy bands of the two instruments and spectral evolution with time \citep{Sakamoto16}.

In this work, pointed observations of \astro\ - SXT and LAXPC have been used to study the timing and spectral variability in \src. Table~\ref{obslog} gives the details of these observations. The \astro\ observation, shown with a dashed blue line in Figure~\ref{fig:longtermlc}, was made during the rising phase of 2021 outburst. For reference, the epoch of the \textit{NICER} observation used by \citet{Mandal22} has been marked with a solid green line. 

SXT consists of a charge coupled device camera operating in the Photon Counting (PC) mode with a time resolution of 2.37 s and in the Fast Windowed (FW) mode having a time resolution of 0.278 s. It is capable of performing X-ray imaging and spectroscopy with an energy resolution of $\sim$150 eV in the 0.3--8.0 keV energy range. The on-axis effective area is $\sim$90 cm$^2$ at 1.5 keV \citep[See,][for details.]{Singh16, Singh17}.

\begin{table*}
\caption{Log of \astro\ observation (Obs Id: A10\_121T01\_9000004134) of \src\ used in this work.}
\centering
\resizebox{1.9\columnwidth}{!}{
\begin{tabular}{l l l l l l l l l}
\hline \hline
\astro      &   Start Time              & Stop Time             & Mode  & Observation   & Net Exposure  &  Mean Count   & Energy        \\
Instrument  &   (yyyy-mm-dd hh:mm:ss)   & (yyyy-mm-dd hh:mm:ss) &       & Span (ks)     & (ks)          &  Rate         & Range (keV)   \\
\hline
SXT         & 2021-01-25 01:48:30       & 2021-01-27 18:36:45   & PC    & 233.3             & 49.8          &   1.68    & 0.7-7.0\\
LAXPC       & 2021-01-25 01:09:07       & 2021-01-27 18:36:44   & EA    & 232.6             & 113.5         &   115     & 3.0-25.0\\
\hline
\end{tabular}}
\label{obslog}
\end{table*}

The level-1 SXT data was processed by using \texttt{sxtpipeline} version 1.4b which generated the filtered level-2 cleaned event files. The \texttt{julia} tool was used to merge the cleaned event files from different orbits in SXT data. Thereafter, \texttt{xselect} v2.5b tool was used to extract the image, light curve and spectra of \src. A circular region of 15 arcmin radius was considered as a source region around the source location. A barycentric correction was applied on the light curve by using the \textsc{ftool}-\texttt{earth2sun}\footnote{\url{https://heasarc.gsfc.nasa.gov/ftools/fhelp/earth2sun.txt}}.
The ancillary response file (ARF) was created with \textsc{sxtARFModule} tool by using the arf file provided by the SXT team. The response file (sxt\_pc\_mat\_g0to12.rmf) and the blank sky background spectrum file (SkyBkg\_comb\_EL3p5\_Cl\_Rd16p0\_v01.pha) provided by the SXT team were used\footnote{\url{https://www.tifr.res.in/~astrosat_sxt/dataanalysis.html}}. 

LAXPC consists of three co-aligned proportional counters (LAXPC10, LAXPC20, LAXPC30) covering an energy range of 3.0--80.0 keV. These detectors have a time resolution of 10 $\mu$s and their combined effective area is about 6000 $\mathrm{cm}^2$. In this work, Event Analysis (EA) mode data extracted from the top layer of LAXPC20 instrument has been used. Data from the other two instruments (LAXPC10 and LAXPC30) were not used due to their unpredictable high-voltage variations and gas leakage \citep[See,][for details.]{Antia17, Antia21}. The level-1 data was processed using the LAXPC data analysis pipeline (\texttt{LaxpcSoft}: version 3.4.4)\footnote{\url{https://www.tifr.res.in/~astrosat_laxpc/LaxpcSoft.html}}. Light curve and spectral files were generated using \texttt{laxpcl1} tool and background products were generated using the \texttt{backshiftv3} command. The photon arrival times were corrected for the solar system barycenter using the \texttt{as1bary}\footnote{\url{http://astrosat-ssc.iucaa.in/?q=data\_and\_analysis}} tool. Since the spin period measurements of a neutron star gets modified due to its orbital motion, therefore, the all barycentered light curves were corrected for orbital motion by using the ephemeris of \citet{Finger96} and \citep{Inam04}.

\begin{figure*}
    \centering
    \includegraphics[width=2.5 in, height=3.0 in, angle=-90]{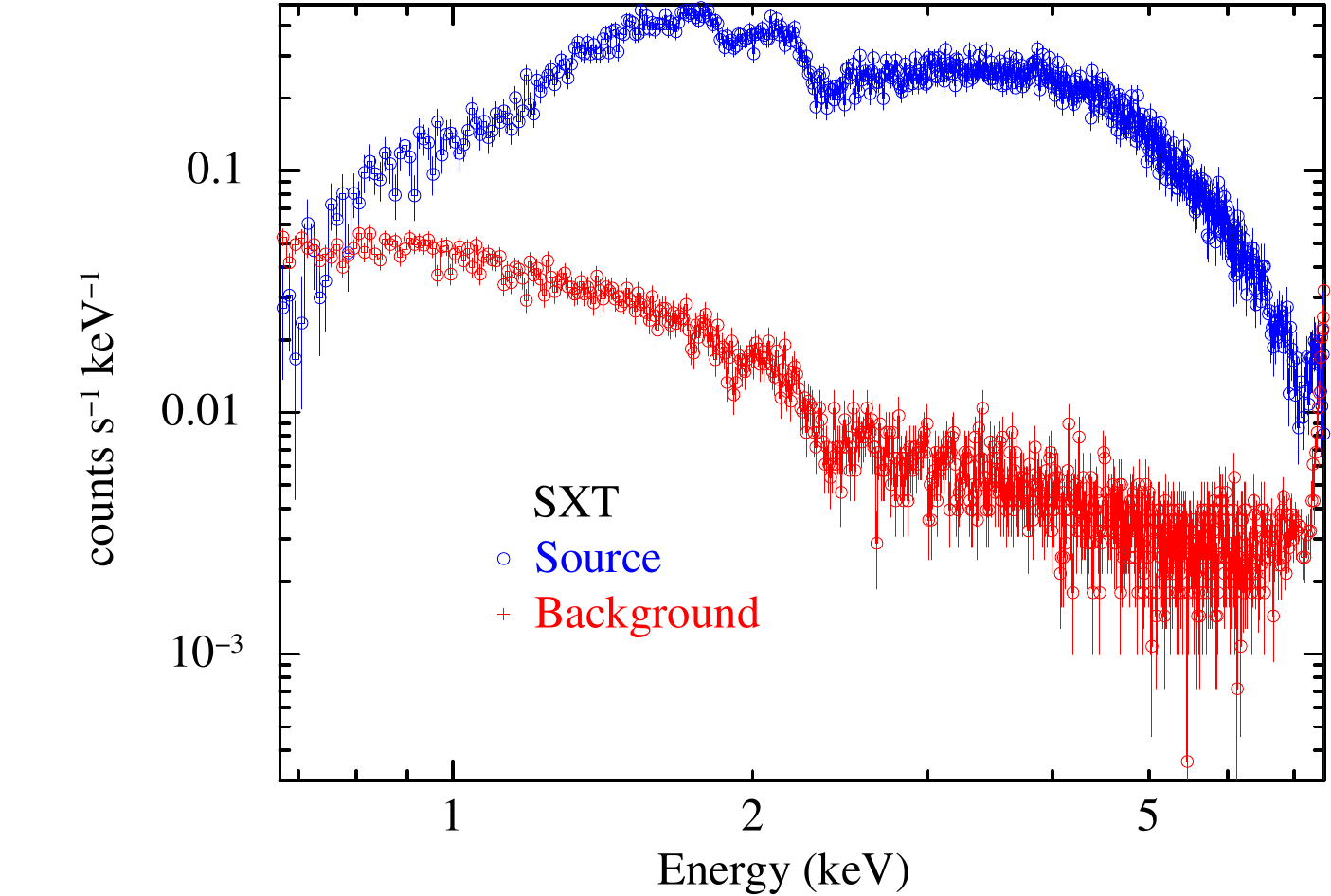}
    \includegraphics[width=2.5 in, height=3.0 in, angle=-90]{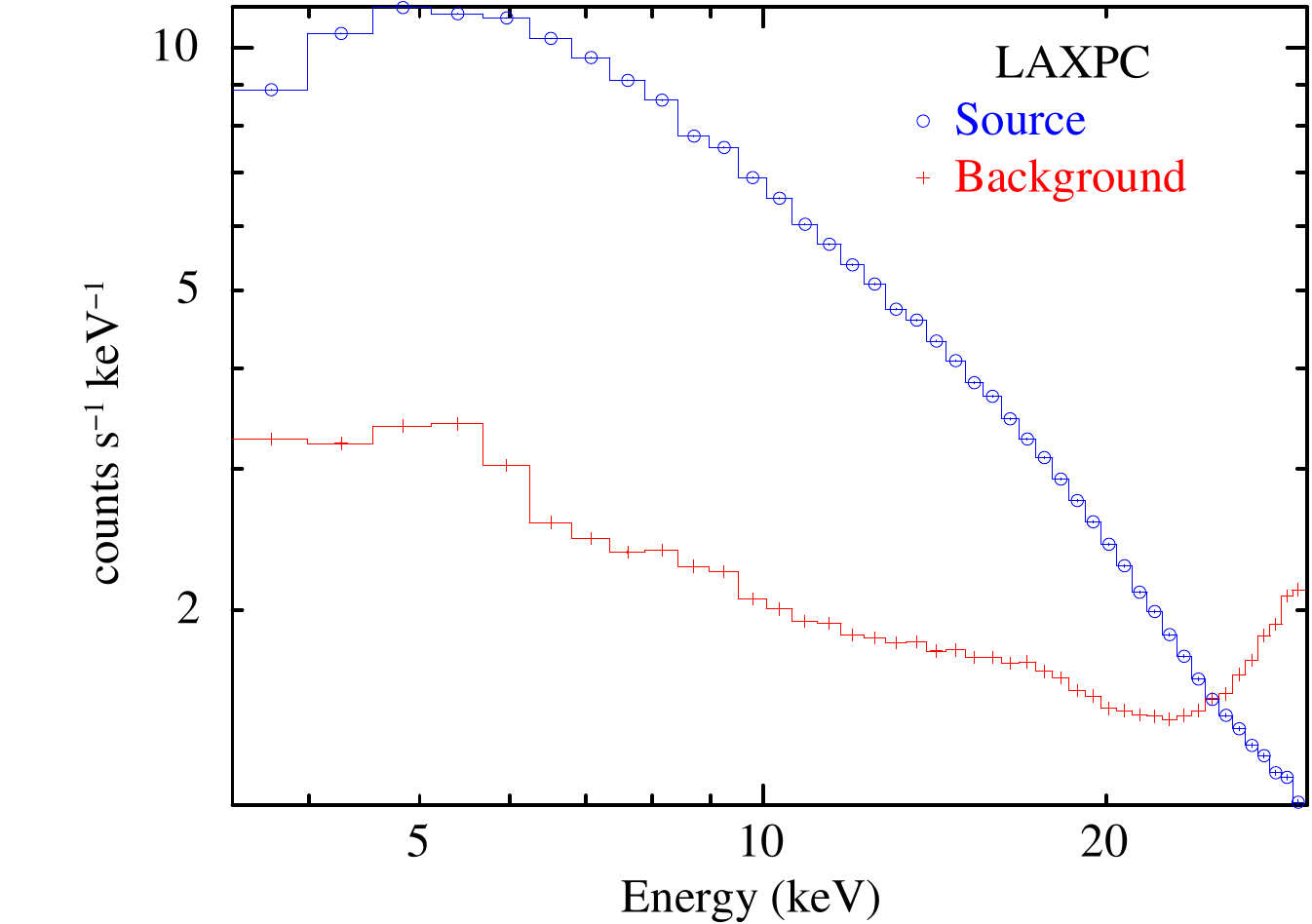}
\caption{The source and background spectrum of \src\ for the \astro\ observation analyzed in this work. The source was visible above the background in the energy range 0.7--7.0 keV and 3.0-25.0 keV, respectively for SXT and LAXPC data.}
\label{fig:srcabovebkg}
\end{figure*}

Figure~\ref{fig:srcabovebkg} shows the source and background spectra of \src\ for both the instruments. During the \astro\ observation, the source was visible above the background in the energy range 0.7--7.0 keV for SXT and 3.0-25.0 keV for LAXPC data. Therefore, the analysis is restricted to these energy ranges only. 

\section{Analysis} \label{sec:analysis}

\begin{figure}
    \centering
    \includegraphics[width=3.5 in, height=3.5 in, ,angle=-90]{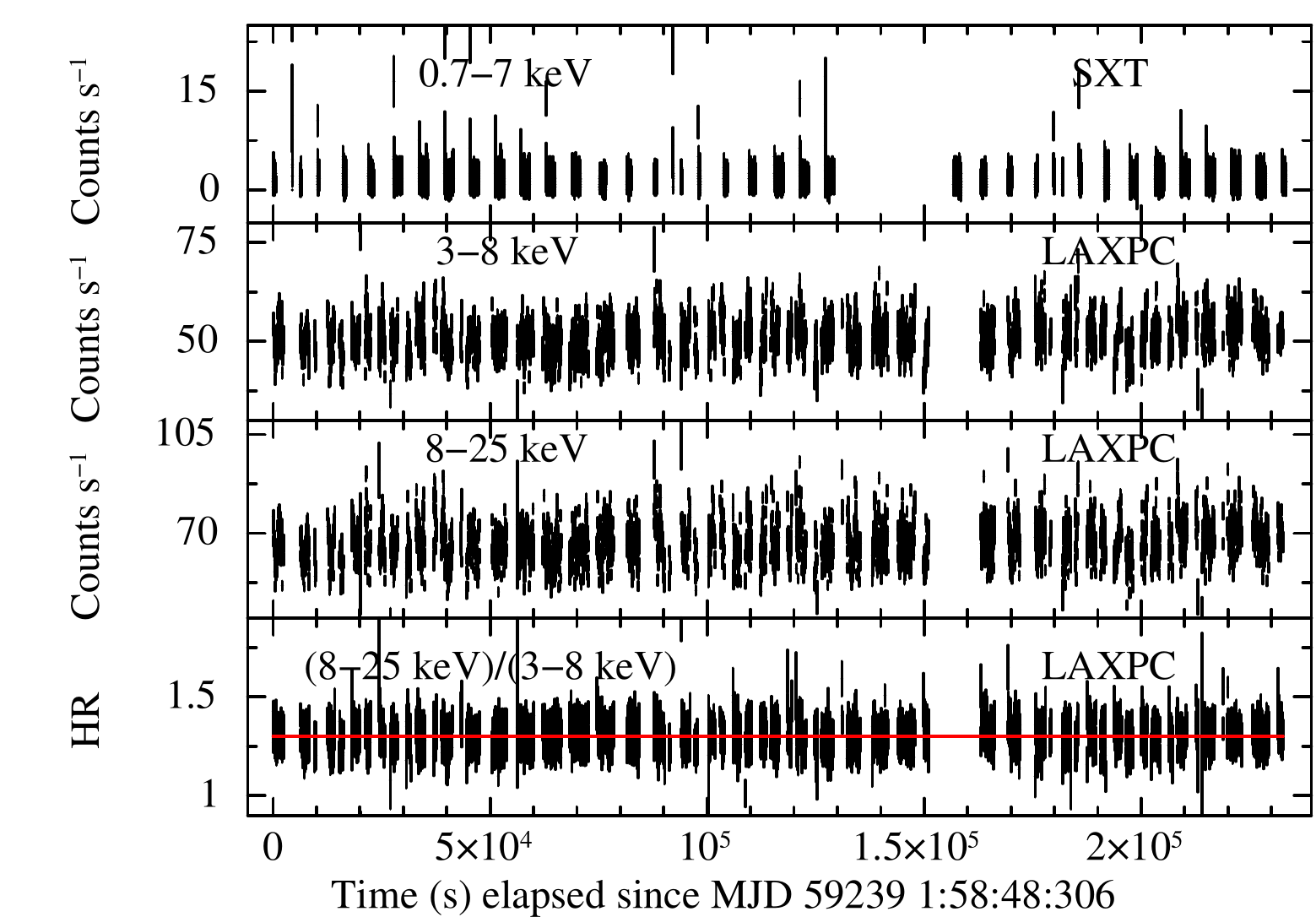}
    \caption{The light curves of \src\ from both the instruments. \textit{Top panel}: The 0.7-7.0 keV SXT light curve. \textit{Second and third panel}: The 3.0--8.0 keV and 8.0--25.0 keV LAXPC light curve. \textit{Bottom panel}: The hardness ratio (HR) plotted as a function of time. The red dash line in the bottom panel corresponds to average ratio of 1.3.}
    \label{fig:hr}
\end{figure}

The background-subtracted SXT and LAXPC light curves of \src\ are shown in Figure~\ref{fig:hr}. The top panel corresponds to 0.7-7.0 keV SXT light curve. The second and third panels in this plot correspond to the 3.0-8.0 and 8.0-25.0 keV LAXPC light curves, respectively. The bottom panel corresponds to the hardness ratio (HR), defined as the ratio of count rate in the hard energy range (8.0-25.0 keV) to that in soft energy range (3.0-8.0 keV). The red dash line in the bottom panel is the reference line at an average HR of 1.3. Clearly, the intensity of \src\ did not vary significantly during the \astro\ observation and the entire data has been used for further analysis.

\subsection{Timing Analysis}

\begin{figure*}
    \centering
    \includegraphics[width=0.3\textwidth,angle=-90]{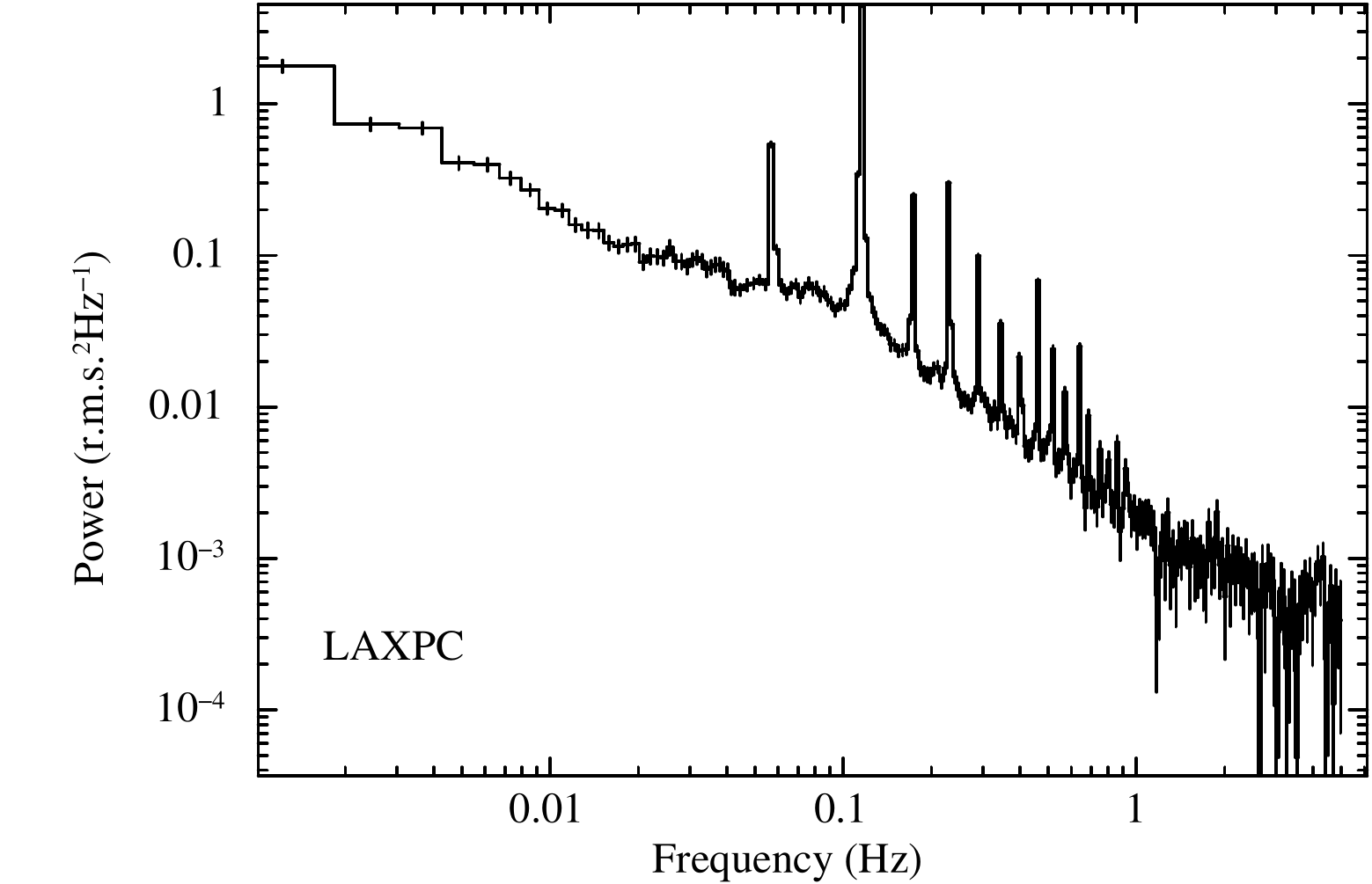}
    \includegraphics[width=0.3\textwidth,angle=-90]{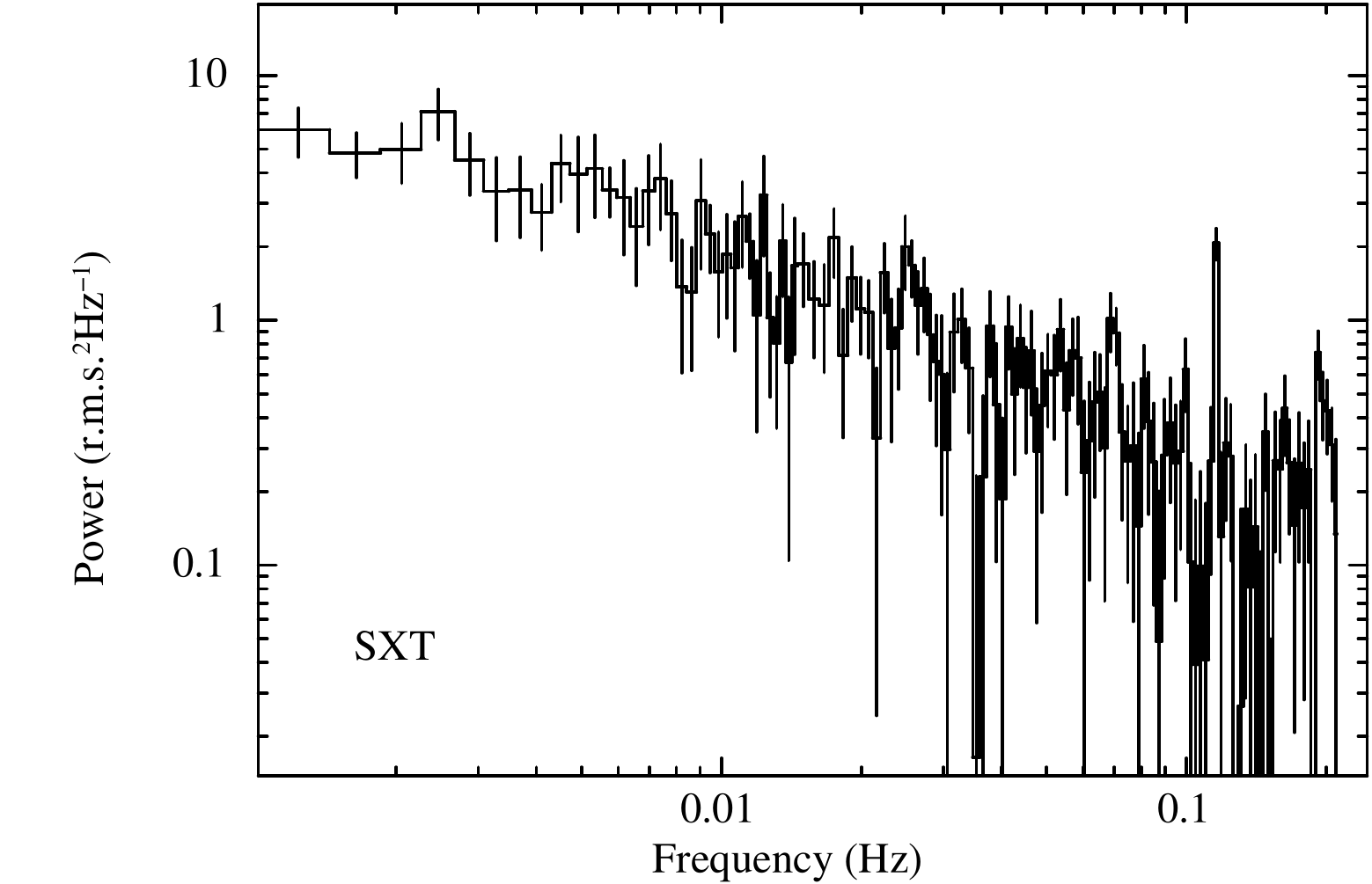}
    \caption{The power density spectrum (PDS) of \src\ obtained from its \astro\ observations. \textit{Left panel}: The PDS obtained from LAXPC data, showing peaks corresponding to fundamental frequency of 0.057 Hz and its harmonics. \textit{Right panel}: The PDS obtained from SXT data did not show any appreciable power above the white noise level. Only the first harmonic is visible at a frequency of 0.114 Hz. }
    \label{fig:pds}
\end{figure*}

Figure~\ref{fig:pds} shows the power density spectrum (PDS) generated from 0.7-7.0 keV SXT (left panel) and 3.0--25.0 keV LAXPC (right panel) light curve of \src\ by using the \texttt{powspec} tool of XRONOS sub-package of \textsc{ftools} \citep{Blackburn99}. The LAXPC light curve binned at 0.1 s was divided into 284 stretches of 8192 bins. However, out of these 284 intervals, 114 intervals were rejected for default window selection due to less data. The PDS were normalised such that their integral gives the squared r.m.s. fractional variability after removal of white noise \citep{Leahy83}. The white noise level for LAXPC data was 0.024. All the PDS were then averaged and re-binned geometrically in frequency. The resultant PDS comprised of a sharp spin frequency peak at around 57 mHz and its harmonics. The fundamental frequency was detected at $\sim$5.5$\sigma$ significance level \citep{Beri23}. 

Owing to SXT's bin time of 2.3775 s and a low count rate during the observation, the SXT light curve was divided into 96 intervals of 1024 bins. As in case of LAXPC data, out of 96 intervals, 59 intervals were rejected due to less data. The PDS generated with SXT light curve did not show any appreciable power above the white noise level of 1.823 and the fundamental frequency at 57 mHz was almost buried in the noise with an r.m.s. value of just $\sim$2\%. Although the first harmonic around 114 mHz was stronger than the fundamental frequency, but its r.m.s. value was also small ($\sim$4\%).  

\begin{figure*}
\centering
\includegraphics[width=2.0 in, height=3.0 in, angle=-90]{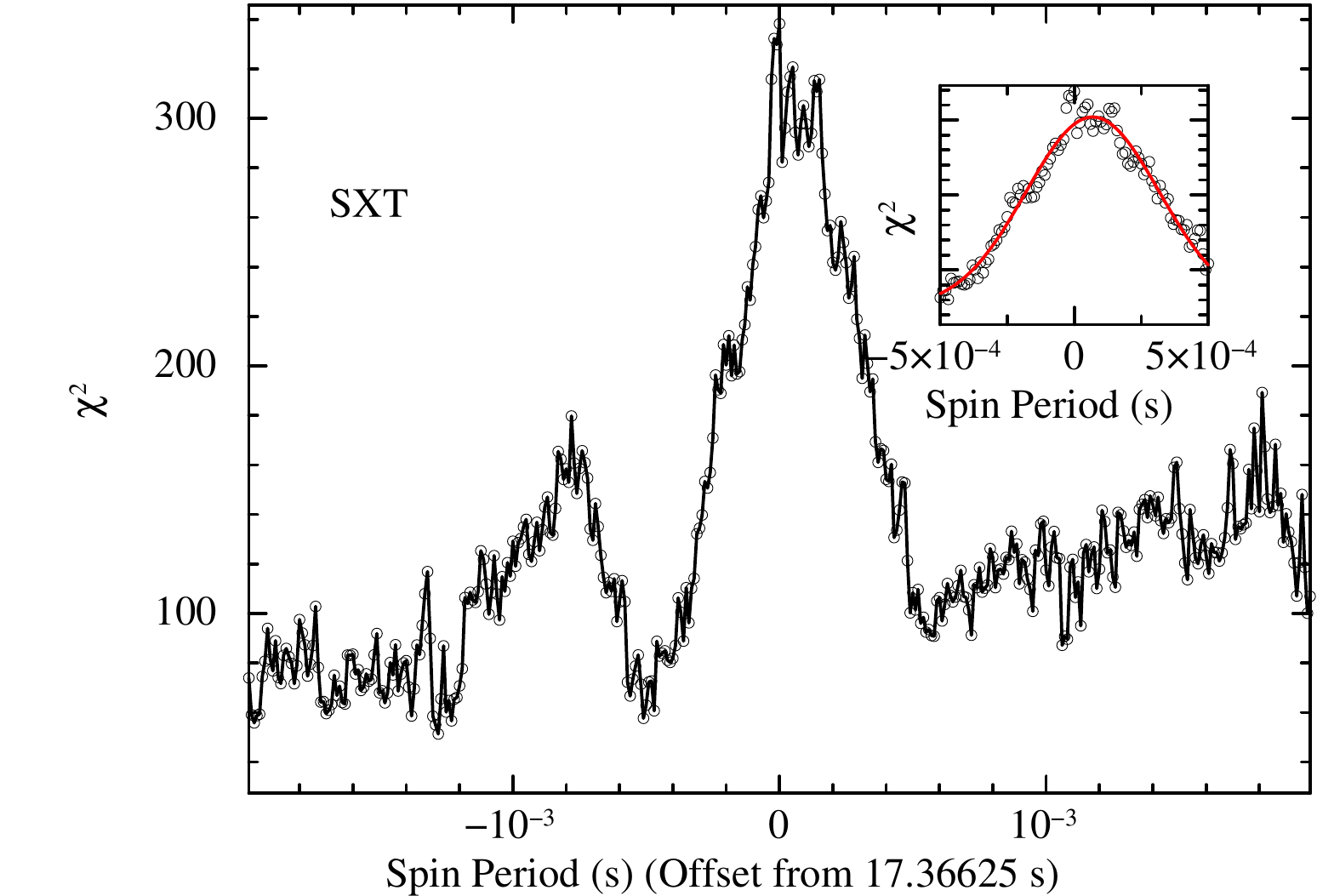}
\includegraphics[width=2.0 in, height=3.0 in, angle=-90]{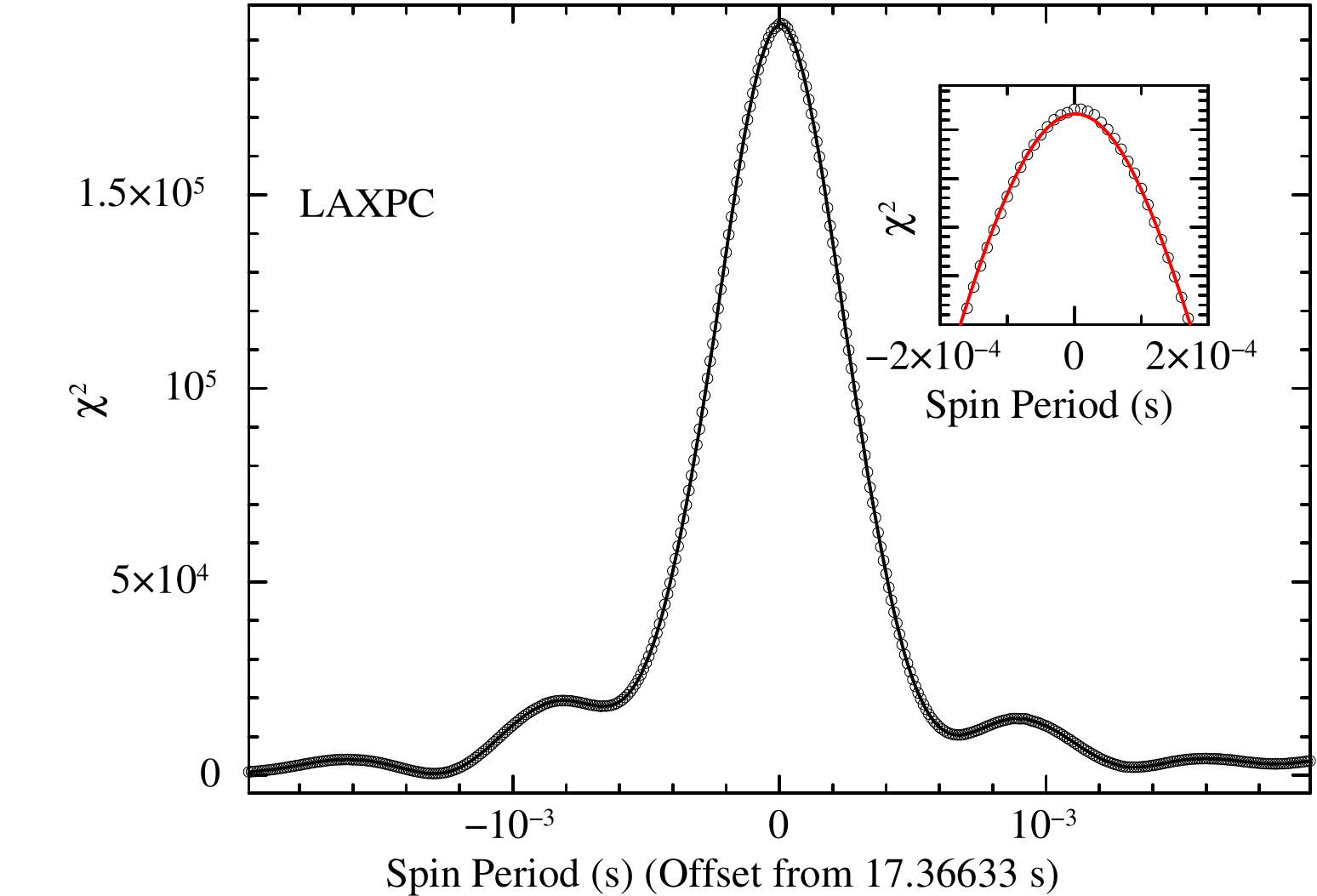}
\caption{The $\chi^2$ variation for different spin periods for \src, in accordance with $\chi^2$ maximisation technique. The \texttt{efsearch} results are shown for SXT data (in left panel) and LAXPC data (in right panel). The inset figures in both the panels show the zoomed in section of the variation near the peak with axes in same units as the main figure. The red curve in the inset figures depict the best fit Gaussian curve.}
\label{efsearch}
\end{figure*}

In order to determine the best estimate of the spin period of neutron star, the \texttt{efsearch} tool of \textsc{FTOOLS}, based on epoch folding and $\chi^2$ maximisation technique \citep{Leahy87} was used. During the \astro\ observation (MJD 59239), the spin period of the neutron star is expected to be around 17.366 s (57.583 mHz at MJD 59240, as measured with Fermi Gamma-Ray Burst Monitor (GBM))\footnote{\url{https://gammaray.nsstc.nasa.gov/gbm/science/pulsars/lightcurves/2s1417.html}}. \citet{Mandal22} reported it around 17.3649 s at MJD 59241 from \textit{NICER} data. Therefore, the SXT and LAXPC time series data, binned at 2.378 s and 0.1 s, respectively, were folded with 400 different trial periods over a narrow range of 17.364 -- 17.368 s with a high resolution of 0.01 ms, and having 32 phase bins in each period. The resultant variation of $\chi^2$ values as a function of trial periods is shown in Figure~\ref{efsearch}. Owing to a low count rate and smaller exposure time during the observation, the $\chi^2$ values of SXT data are much less than that in the LAXPC results \citep{Jain09}. The $\chi^2$ variation was fit with a Gaussian profile to determine the period corresponding to the maximum $\chi^2$ (shown as inset in this figure with axis in same units). The best fit model gave a spin period of 17.36632 s for SXT data and 17.366332 s for LAXPC data. The bootstrap method of \citet{Boldin13} was used to determine the error in spin period. Following the method described in \citet{Sharma23}, 1000 light curves were simulated and spin period was determined from each of them by using epoch folding technique. The standard deviation in the best period determination was taken as the error in the pulse period. As a result, the best pulse period of \src\ was found to be 17.36632(2) s (SXT) and 17.366332(5) s (LAXPC) at MJD 59239.082. These estimated values of the pulse period are in good agreement with the orbitally-corrected frequency history reported by Fermi/GBM team near the epoch of \astro\ observation.

\begin{figure*}
\centering
\includegraphics[width=3.5 in, height=3.0 in, angle=-90]{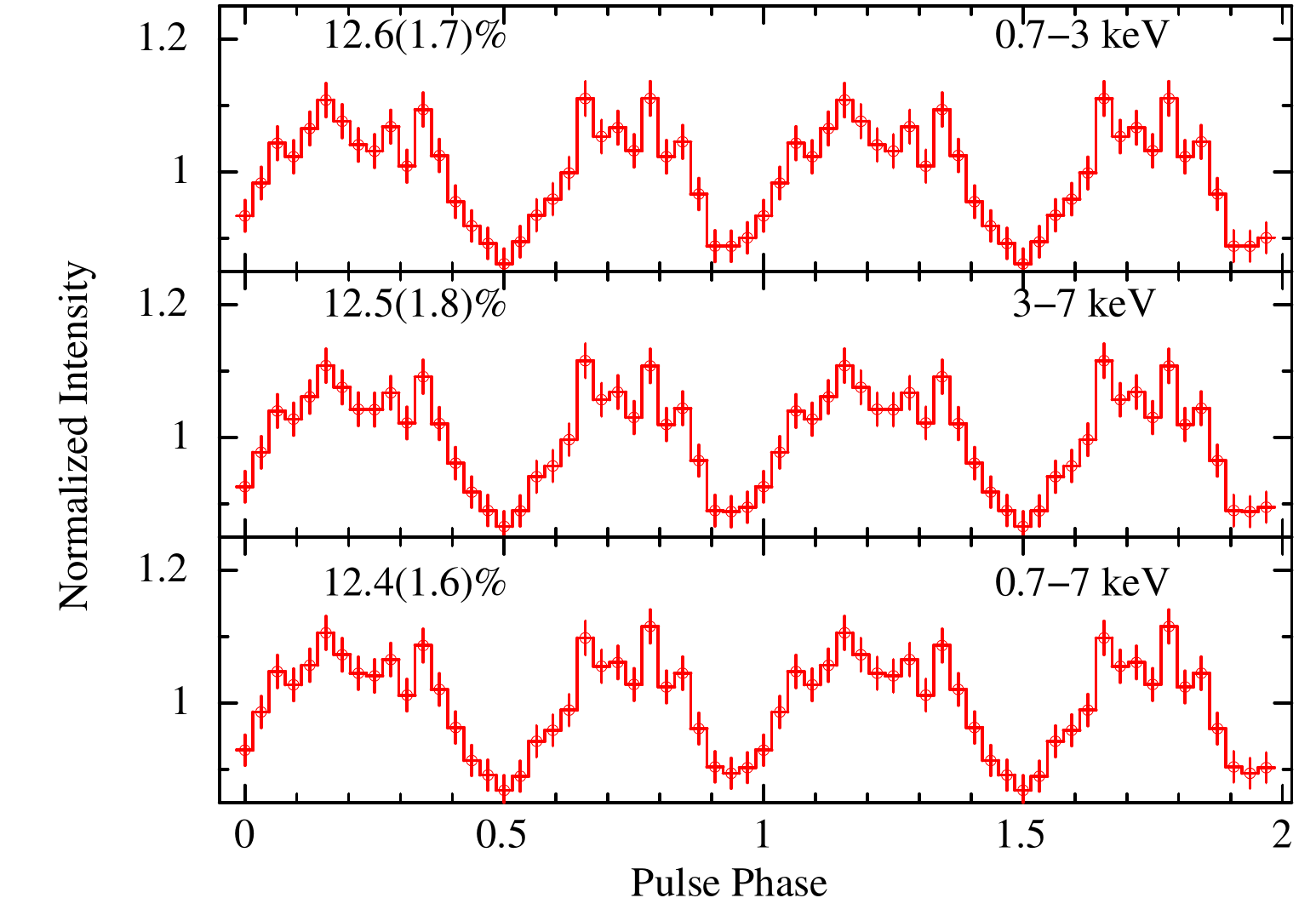}
\includegraphics[width=3.5 in, height=3.0 in, angle=-90]{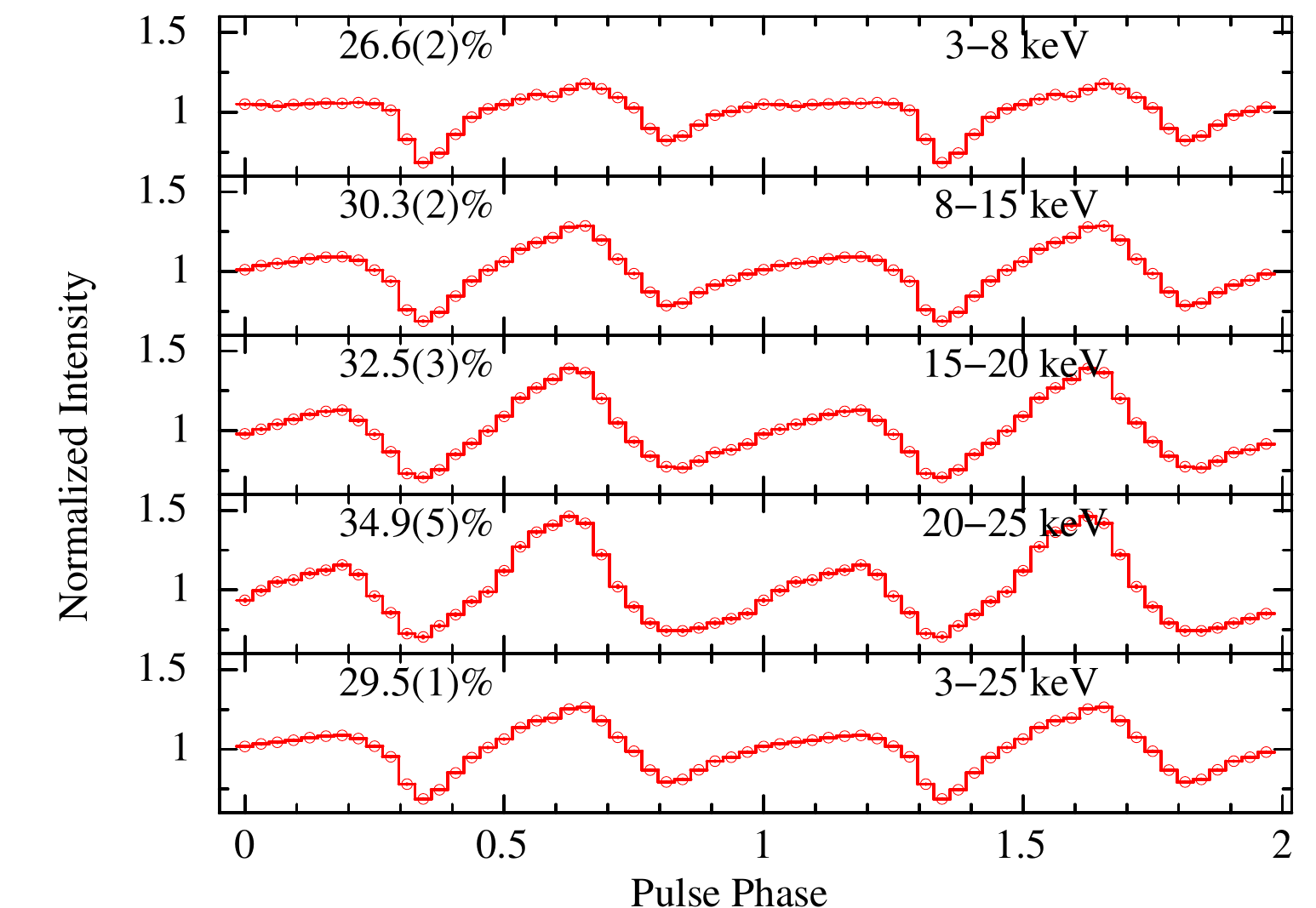}
\caption{Energy resolved pulse profile of \src. \textit{Left panel}: The SXT profiles in 0.7-3.0, 3.0-7.0 and 0.7-7.0 keV energy range. \textit{Right panel}: The LAXPC profiles in 3.0--8.0, 8.0--15.0, 15.0--20.0, 20.0--25.0 and 3.0--25.0 keV energy range. Two cycles of pulse phase have been shown for clarity. The pulsed fraction and energy range is mentioned in each panel.}
\label{fig:efold}
\end{figure*}

Figure~\ref{fig:efold} shows the energy resolved pulse profiles of \src\ in the energy range of 0.7-3.0, 3.0-7.0, 0.7-7.0 keV for SXT and 3.0--8.0, 8.0--15.0, 15.0--20.0, 20.0--25.0 and 3.0--25.0 keV for LAXPC, generated using \texttt{efold} tool of \textsc{FTOOLS}. The pulse profiles were found to be strongly dependent on energy. The profile consists of two peaks at all energies. The low energy peaks in the SXT profiles were broad and comprised of several mini-structures. The pulsed fraction\footnote{$\left(\frac{I_{max}-I_{min}}{I_{max}+I_{min}}\right)*100$, where I$_{max}$ and I$_{min}$ are respectively, the maximum and minimum intensity in the folded light curve.} was low compared to LAXPC profiles. In the LAXPC profiles, the relative intensity of main peak was observed to increase with energy. Both the peaks were broad at low energies and sharpened with energy. The pulsed fraction increased significantly with energy.

\subsection{Spectral Analysis}

The combined broadband 0.7--25.0 keV spectral modelling of \src\ with spectra from SXT and LAXPC has been performed using \texttt{xspec} version 12.14.0b from \textsc{Heasoft} 6.33.1 package \citep{Arnaud96}. The WILM abundances was used to model the absorption of X-rays in the interstellar medium \citep{Wilms00} and photo-ionization cross sections of \citet{Verner96} for the spectral modelling. The SXT and LAXPC spectra were grouped using \texttt{grppha} to have a minimum count of 20 counts per bin. A cross-calibration constant was added for the two instruments. It was fixed to 1 for LAXPC and was allowed to vary for SXT. A 3\% systematic uncertainty was used during spectral fitting as prescribed by the POC \citep{Antia17, Bhattacharya17, Antia21}. A gain correction was also applied for the SXT spectrum. The gain offset of response matrix was allowed to vary with its slope frozen to 1. For each parameter, the \texttt{err} command was used to determine the uncertainty in its estimated value. All the spectral uncertainties reported in this work are at a 90\% confidence level ($\Delta \chi^2$ = 2.7). 

\begin{figure*}
\centering
\includegraphics[width=0.7\columnwidth,angle=-90]{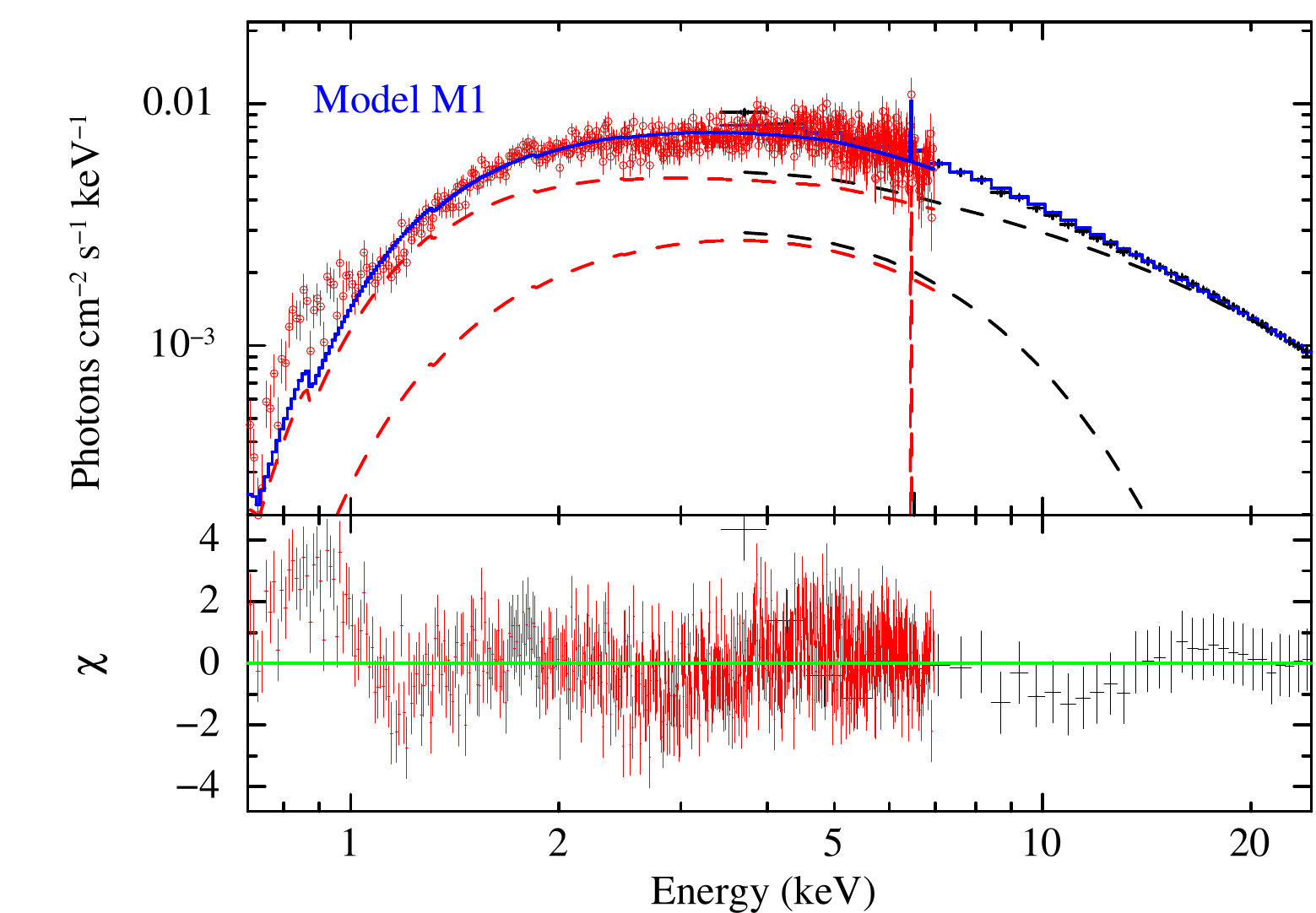}
\includegraphics[width=0.7\columnwidth,angle=-90]{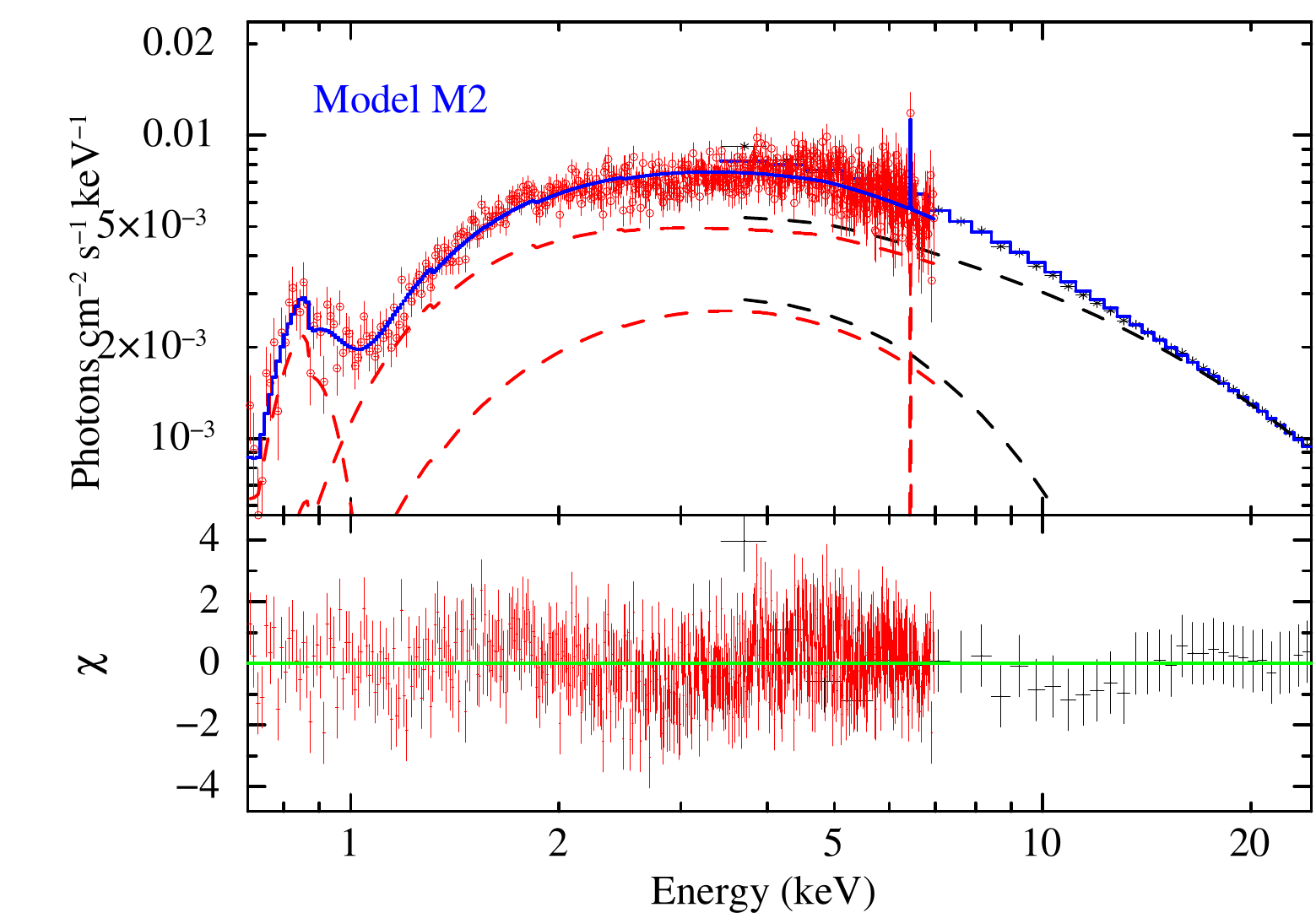}
\includegraphics[width=0.7\columnwidth,angle=-90]{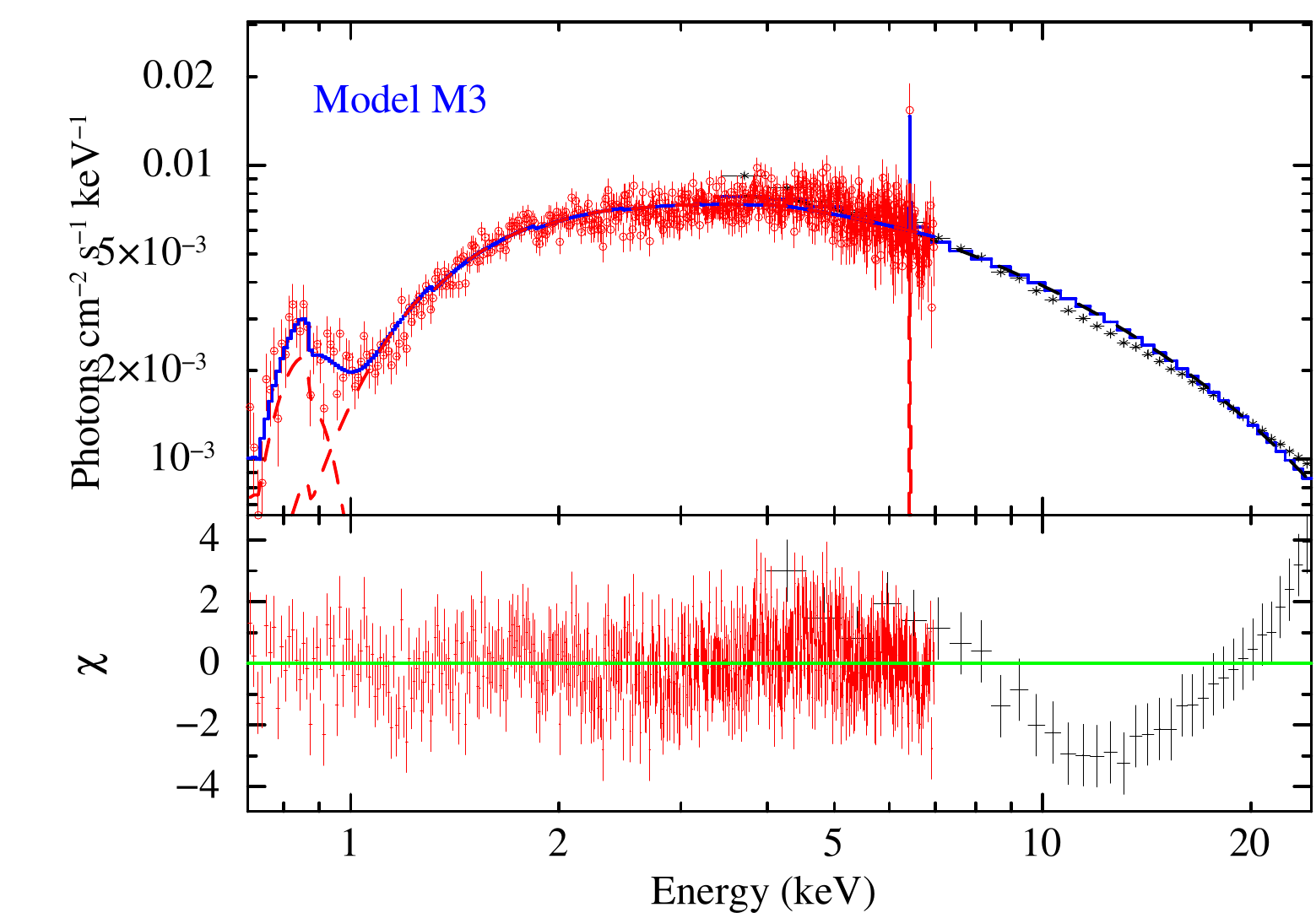}
\includegraphics[width=0.7\columnwidth,angle=-90]{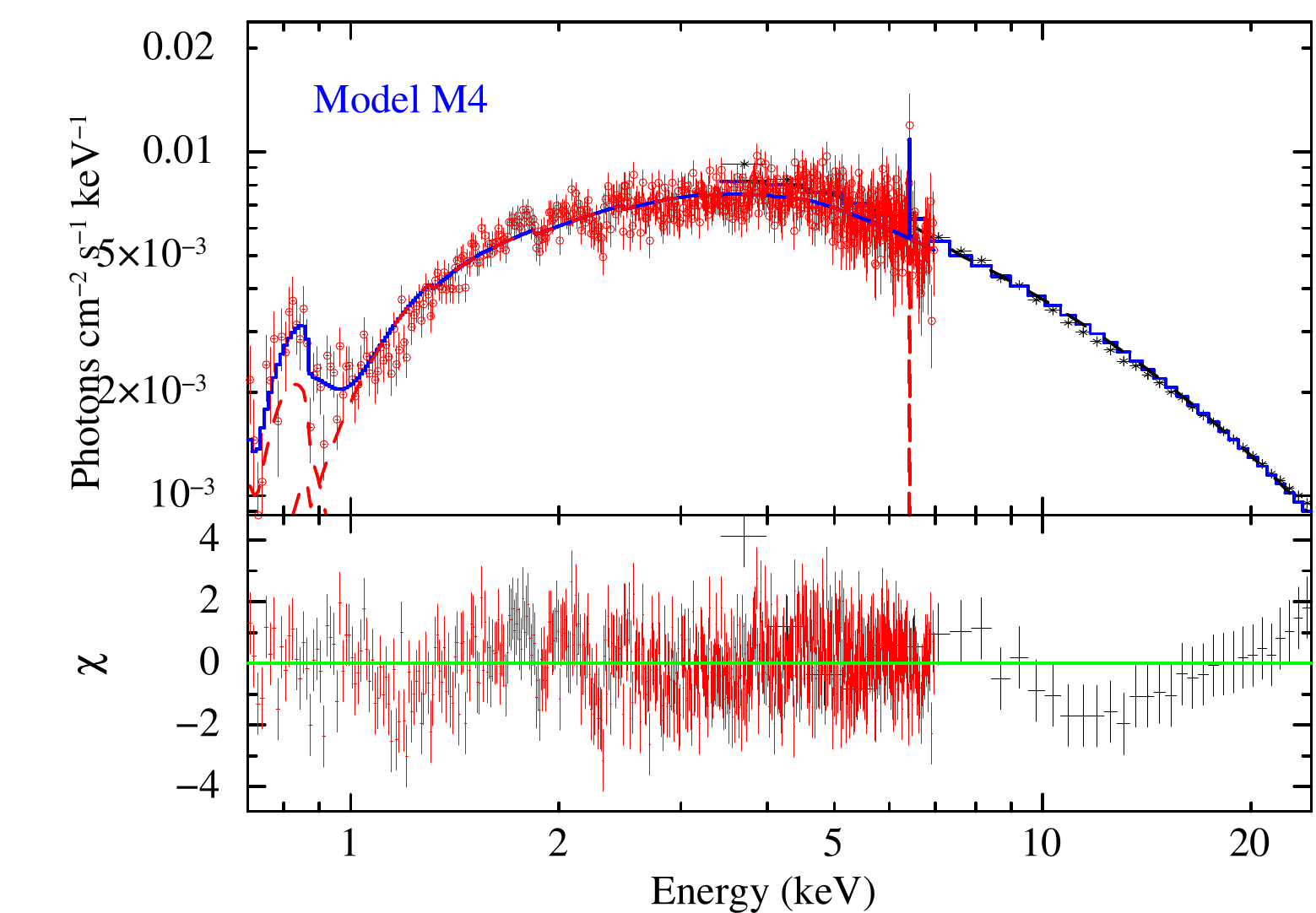}
\caption{The 0.7--25.0 keV unfolded energy spectrum of \src, generated from simultaneous fitting of SXT (0.7-7.0 keV: Shown in red) and LAXPC (3.0-25.0 keV: Shown in black) data. The best fit models are shown with a solid blue line. \textit{Top-Left graph}: Model M1 (\texttt{constant*phabs*(powerlaw*highEcut+bbodyrad+gaussian)}). \textit{Top-Right graph}: Model M2 (\texttt{constant*phabs*(powerlaw*highEcut+bbodyrad+gaussian+gaussian)}). \textit{Bottom-Left graph}: Model M3 (\texttt{constant*phabs*(powerlaw*highEcut+gaussian+gaussian)}).\textit{Bottom-Right graph}: Model M4 (\texttt{constant*phabs*pcfabs*(powerlaw*highEcut+gaussian+gaussian)}). The dashed lines in all the graphs reflect the individual spectral components of respective model. The bottom panel of all the four graphs show the residual from the respective best fit model.}
\label{fig:spec}
\end{figure*}

Following \citet{Mandal22}, the 0.7--25.0 keV energy spectrum of \src\ was initially fit with model comprising of high energy cut-off power law (\texttt{powerlaw*highEcut}) along with photoelectric absorption (\texttt{phabs}), blackbody component (\texttt{bbodyrad}) to account for soft excess and a Gaussian profile (\texttt{gaussian}) for the emission feature around 6.5 keV (hereafter Model \textit{M1}). The equivalent hydrogen column density (N$_H$) of photoelectric absorption component was fixed to the Galactic survey value\footnote{\url{https://heasarc.gsfc.nasa.gov/cgi-bin/Tools/xraybg/xraybg.pl}} of 1.12$\times$ 10$^{22}$ cm$^{-2}$ \citep{nh16}. The best fit for this model had a $\chi^2$ of 773 for 619 degrees of freedom (d.o.f.) ($\chi^2_{red}$ 1.25). Figure~\ref{fig:spec} shows all the models used for fitting the \src\ spectra. The respective best fit parameters are listed in Table~\ref{tab:spec}. 

\begin{figure*}
    \centering
    \includegraphics[width=\columnwidth]{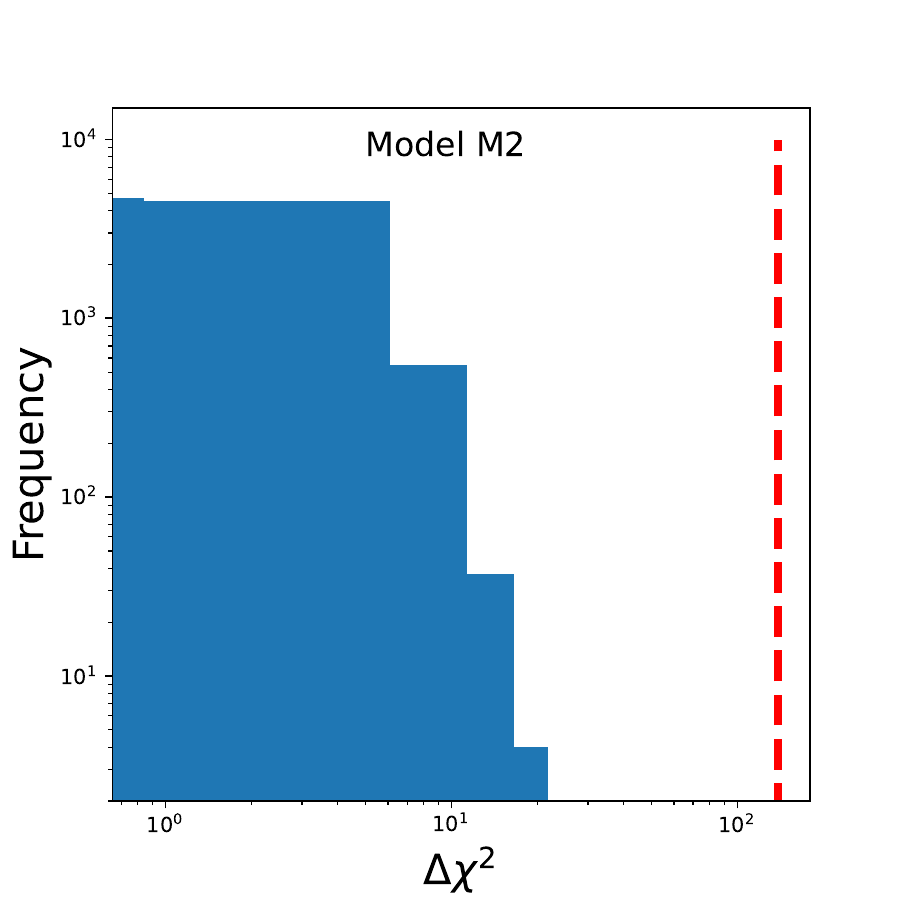}
    \includegraphics[width=\columnwidth]{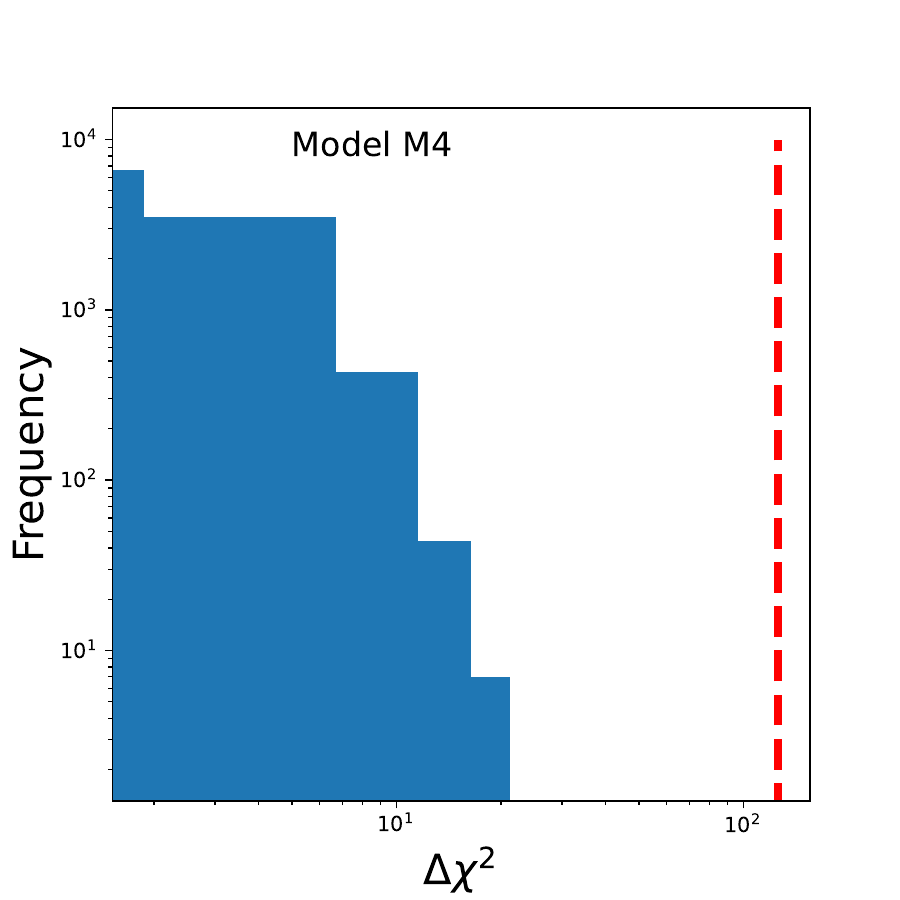}
    \caption{The histogram showing the result of \texttt{simftest} simulations for estimating the significance of 0.83 keV emission line in the energy spectrum of \src\ for models \textit{M2} (left panel) and \textit{M4} (right panel). The vertical line in both the panels corresponds to observed $\Delta \chi^2$ of 139 and 126, respectively.}
    \label{fig:simftest}
\end{figure*}

In order to account for the excess in residual around 0.8 keV in best fit Model \textit{M1}, additional Gaussian component was added (hereafter Model \textit{M2}). This improved the fit significantly with $\Delta\chi^2$ of 139 for 3 additional d.o.f. This is the first time that 0.83 keV emission line has been detected in \src. In order to check the statistical significance of presence of this emission line, the \texttt{simftest} routine of \texttt{xspec} was used. This routine uses Monte Carlo simulations to generate simulated spectra based on the real observed spectra and evaluates the difference in $\Delta \chi^2$ for an additional model component, which in this case is a Gaussian emission feature. Using \texttt{simftest}, 10,000 spectra were generated. The histogram of Figure~\ref{fig:simftest} shows the simulation results, which clearly indicates a large difference between the observed $\Delta\chi^2$ (139) and the maximum value (22) estimated from the simulations. The detection significance is more than 3$\sigma$ confidence level. It is necessary to mention here that the spectral analysis of \src\ with data from other X-ray missions has not detected the presence of 0.83 keV spectral line. Therefore, even though thorough statistical checks support the presence of this line, the residuals around 0.83 keV could be an instrumental effect. 

A black body temperature of 1.97 keV and 1.61 keV was obtained from models \textit{M1} and \textit{M2}, respectively. Although a high black body temperature is not very unusual in transient Be X-ray binaries (example, GRO J1008-57) \citep{Shrader99}, but it is relatively large compared to $\sim$0.255 keV (\citet{Mandal22}), $\sim$0.56 keV (\citet{Serim22}) and $\sim$0.9 keV (\citet{Gupta19}) reported in previous works. Similarly, in both these models, the spectral index was about 0.28, which is quite small compared to $\sim$0.41 reported in previous works. Therefore, in order to bring out the best spectral-fit, more models without inclusion of black body component were used to fit the energy spectrum of \src. The first model, (hereafter, Model \textit{M3}) was similar to \textit{M2}, except the black body component. The best fit \textit{M3} model had a higher $\chi^2$ compared to model \textit{M2} and there was no significant change in the spectral index. Another model, (hereafter, Model \textit{M4}) comprising of an absorbed power law with a high energy cutoff plus a partial covering absorber (\texttt{pcfabs}) and Gaussian emission lines was also tried. The best fit for this model had a $\chi^2_{red}$ value comparable to that of model M2, spectral index increased to $\sim$0.53 and the best fit had slightly more residual towards higher energies. Since, model \textit{M4} gave equally good spectral fit as model \textit{M2}, therefore, the \texttt{simftest} simulation for detection significance of 0.83 keV emission line was performed for model M4 also. Just like the case of model \textit{M2}, histogram in Figure~\ref{fig:simftest} shows a large difference between the observed $\Delta\chi^2$ (126) and the maximum value (21) estimated from the simulations.

\begin{table*}
	\centering
    \caption{The best fit spectral parameters for all the four models used for simultaneous fitting of 0.7-7.0 keV SXT and 3.0-25.0 keV LAXPC observations of the \src\ spectra. All the errors quoted are within 90 \% confidence range ($\Delta \chi^2 = 2.7$).}
    \label{tab:spec}
    \begin{tabular} {l l l l l l}
    \hline\hline
    Model               & Parameters                    &   \multicolumn{4}{c}{Spectral Model}\\
    \cmidrule[\heavyrulewidth](l){3-6}

    Component           & Value                         &   \textit{M1}$^{a}$       & \textit{M2}$^{b}$         &   \textit{M3}$^{c}$       &   \textit{M4}$^{d}$           \\
    \hline
    \texttt{phabs}      & N$_H$ (10$^{22}$ cm$^{-2}$)   &   1.12$^{fixed}$          &   1.12$^{fixed}$          &   1.12$^{fixed}$          &   1.12$^{fixed}$              \\[1ex]

    \texttt{pcfabs}     & N$_H$ (10$^{22}$ cm$^{-2}$)   &   --                      &   --                      &   --                      &   2.3$_{-1.4}^{+1.3}$          \\
                        & Covering Fraction             &   --                      &   --                      &   --                      &   0.42$_{-0.06}^{+0.17}$       \\[1ex]
    
    \texttt{bbodyrad}   & kT(keV)                       &   1.97$\pm0.12$           &   1.61$\pm0.12$           &   --                      &   --                          \\
                        & norm$_{BB}$(10$^{-3}$)        &   2.35$_{-0.25}^{+0.33}$    &   2.18$_{-0.38}^{+0.44}$    &   --                  &   --                          \\[1ex]
                        
    \texttt{powerlaw}   & $\Gamma$                      &   0.29$\pm0.05$           &   0.28$\pm0.05$           &   0.21$\pm0.02$           &   0.53$\pm0.08$               \\
                        & norm(10$^{-3}$)               &   7.9$\pm0.4$             &   8.2$\pm0.4$             &   11.3$\pm0.3$            &   16.0$\pm2.0$             \\[1ex]
                        
    \texttt{highecut}   & E$_{cutoff}$(keV)             &   4.92$_{-0.35}^{+0.45}$  &  4.91$_{-0.32}^{+0.43}$   &   4.2$_{-0.3}^{+0.2}$     &  4.87$\pm0.3$                 \\
                        & E$_{fold}$(keV)               &   15.8$_{-1.8}^{+2.7}$    &   15.1$_{-1.4}^{+2.0}$    &   10.1$\pm0.2$            &   13.2$\pm1.1$                \\[1ex]
                        
    \texttt{gaussian}   & E(keV)                        &   6.46$\pm0.07$           &   6.47$\pm0.07$           &   6.46$_{-0.08}^{+0.06}$  &   6.47$\pm0.07$               \\
                        & $\sigma$(keV)                 &   0.005$^{fixed}$         &   0.005$^{fixed}$         &   0.005$^{fixed}$         &   0.005$^{fixed}$             \\
                        & Eqw (eV)                      &   28$_{-18}^{+17}$        &   27$_{-16}^{+19}$        &   34$\pm22$               &   27$_{-16}^{+19}$              \\
                        & norm (10$^{-4}$)              &   1.4$\pm1.2$             &   1.4$\pm1.2$             &   2.1$\pm1.3$             &   2.6$\pm1.2$                 \\[1ex]
                                                                                               
    \texttt{gaussian}   & E(keV)                        &   --                      &   0.83$_{-0.05}^{+0.03}$  &   0.84$_{-0.04}^{+0.03}$  &   0.84$_{-0.05}^{+0.03}$       \\
                        &$\sigma$(keV)                  &   --                      &   0.09$_{-0.02}^{+0.03}$  &   0.07$\pm0.02$           &   0.09$_{-0.02}^{+0.03}$       \\
                        & EQW (eV)                      &   --                      &   687$_{-283}^{+10}$      &   404$_{-92}^{+99}$       &   687$_{-283}^{+9}$           \\
                        & norm (10$^{-3}$)              &   --                      &   6.4$_{-1.6}^{+3.1}$     &   4.56$_{-1.1}^{+1.9}$     &   6.4$_{-1.6}^{+3.1}$          \\[1ex]
                                                                                         
    \texttt{constant}   & C$_{LAXPC}$                   &   1$^{fixed}$             &   1$^{fixed}$             &   1$^{fixed}$             &   1$^{fixed}$                 \\
                        &C$_{SXT}$                      &   0.93$\pm0.02$           &   0.93$\pm0.05$           &   0.94$\pm0.02$           &   0.93$\pm0.05$               \\[1ex]
      Unabs. Flux$^{e}$ & $F_{0.7-25.0 \rm ~keV}^{\rm total}$& 5.36                 &   5.41                    &   6.03                    &   5.41                        \\[1ex]                   

      Gain offset       & eV                            &   12                      &   28                      &   37                      &   42                              \\
                    &$\chi^2$ (d.o.f.)      &   773 (619)           &   634 (616)   &   749 (618)               &   655 (616)          \\
                    &$\chi^2_{red}$       &   1.25           &   1.03   &   1.21               &   1.06          \\
                    
      \hline
    \multicolumn{5}{l}{$^{a}$\texttt{constant*phabs*(powerlaw*highEcut+bbodyrad+gaussian)}}\\
    \multicolumn{5}{l}{$^{b}$\texttt{constant*phabs*(powerlaw*highEcut+bbodyrad+gaussian+gaussian)}}\\
    \multicolumn{5}{l}{$^{c}$\texttt{constant*phabs*(powerlaw*highEcut+gaussian+gaussian)}}\\
    \multicolumn{5}{l}{$^{d}$\texttt{constant*phabs*pcfabs*(powerlaw*highEcut+gaussian+gaussian)}}\\
    \multicolumn{5}{l}{$^{e}$Unabsorbed Flux in the units of $\times 10^{-10}$\erg.}\\

    \end{tabular}
 \end{table*}

\section{Discussion and Conclusion}
\label{sec:discuss}

In this work, the timing and spectral properties of the Be transient X-ray pulsar, \src, during its 2021 outburst have been studied by using data from SXT and LAXPC on-board the \astro. This is the first time that the 2021 outburst of this source has been studied over a broad energy range of 0.7-25.0 keV. The only other known measurements during this outburst are with \textit{NICER} data, with timing analysis in 0.4-10 keV and spectral analysis in 0.8-12.0 keV energy range \citep{Mandal22}.

The main results from the current work include,
\begin{enumerate}
    \item Detection of 17.36632(2) s pulsations in SXT (0.7-7 keV) and 17.366332(5) s in LAXPC (3.0-25.0 keV) data at MJD 59239.082. 
    \item Significant energy dependence of pulse profiles with pulsed fraction increasing with energy.
    \item The 0.7--25.0 keV energy spectrum is well described by a power-law with an exponential cut-off model. A black body component was required in addition to high column density to model the soft excess along with Gaussian emission line feature. Another model comprising of partial covering absorber instead of black body component gives an equally good spectral fit.
    \item The first ever possible detection of an unresolved Fe\texttt{XVII} - Fe\texttt{XIX} line complex at 0.83 keV.
\end{enumerate}

The pulse profiles of both persistent and transient HMXBs are known to show striking variations with energy as well as source flux \citep{White83}. For disk-fed persistent HMXBs such as Cen X-3, shielding of the emission region (hotspots at the magnetic poles) by the plasma layer results in changes in shape of the pulse profile \citep{Basko76}. In systems like 4U 1538-52 and Vela X-1, scattering atmosphere surrounding the accretion column results in energy dependent profiles \citep{Sturner94}. In transient Be-HMXBs, such as, A 1118–61 \citep{Devasia11}, 4U 0115+63 \citep{Tsygankov07}, GX 304-1 \citep{Devasia11b} and EXO 2030+375 \citep{Parmar89}, these variations are presumed to occur due to variable mass accretion rates during outbursts, which results in geometrical variations in the accretion column, thereby affecting the shape of the pulse profiles. Compared to the rising and decay phase of the outburst, the profiles in these systems are relatively more complex near the peak of the outburst. Some Be-HMXB pulsars like A 0535+262 have remarkably different pulse profiles during quiescent and outburst phase \citep{Mukherjee5, Caballero07}.

The pulse profiles of \src\ have shown interesting trend since its first detection in 1978 during which the 2.0-11.0 keV pulse profile showed sharp features along with an inter-pulse structure \citep{Kelley81}. \citet{Finger96} observed double peaked morphology with an off-pulse in between in the 20.0-50.0 keV BATSE data. During the 1999 outburst, the profile was dual peaked and it evolved with the outburst phase \citep{Inam04}. Similarly, \citet{Gupta18} reported a double peaked structure at low luminosity evolving into triple peaked profile near the peak of the outburst and reverting to dual peaked morphology during the decay of the 2009 outburst. These changes were attributed to changes in beam pattern of the pulsar. Again, during the 2018 outburst, the pulse profile evolved dramatically with the outburst phase. The 0.2--10.0 keV profile was double peaked at low flux phase, evolved into triple peak as flux increased and was quadruple peaked at even higher luminosity \citep{Ji20}. The pulse shape evolved from four peaked profile at lower energies into double peaked profile at higher energies \citep{Gupta19}. 

In contrast to the previous outbursts, the only other reported pulse morphology during the 2021 outburst is with \textit{NICER} observations (in effectively 0.4-6.0 keV energy range because of low count rate at higher energies) and it indicated multiple broad peaks and narrow dips \citep{Mandal22}. The \astro\ data taken during the rising part of the outburst (just before the \textit{NICER} observation), therefore provides a good opportunity to study the coherent emission over a broad energy range of 0.7--25.0 keV. 

The SXT-LAXPC profiles were double peaked at all energies with a phase separation of $\sim$0.5. Just like \citet{Mandal22}, both the peaks in the SXT profiles were broad, their relative intensity was similar and they comprised of several mini-structures. There appeared a dip-like feature around 0.9 pulse phase but it was not as resolved as with the \textit{NICER} data. The dip like feature at low energies could be due to photo-electric absorption of soft X-rays photons by the material which is phase-locked to the neutron star \citep{Jaisawal16}. The LAXPC profiles were similar to 2.0--60.0 keV \textit{RXTE} profiles observed during the rising phase of 2009 outburst \citep{Gupta18}. The relative intensity of main peak increased with energy, the broad secondary peak sharpened with increasing energy and the dip like feature was more pronounced at low energies. 

The pulse morphology crucially depends on the extent of interaction of accreting material and geometry of the emitting region. This in turn depends on the mass accretion rate and luminosity of the pulsar \citep{Basko76}. In accretion powered pulsars, emission is mostly due to Comptonization of seed photons located on the surface of neutron star \citep{Becker07}. Emission at sub-critical flux levels, occurs from hot spots at the magnetic poles of neutron star and a pencil beam pattern is expected. But at super-critical levels, a radiation dominated shock front appears in the accretion column due to bulk and a complex beam shape (mixture of pencil and fan beam) is expected.

In coherence with other Be X-ray binary pulsars, such as 4U 0115+63 \citep{Tsygankov07}, EXO 2030+375 \citep{Klochkov08} and Swift J0243.6+6124 \citep{Jaisawal18}, the pulsed fraction of \src\ increases monotonically with energy \citep{Lutovinov09}, without any local features such as those near the cyclotron line energy. It indicates that a large fraction of high-energy photons show pulsating nature in this pulsar. This is often attributed to the fact that the regions emitting hard photons are closer to the neutron star surface and are therefore more compact than the higher regions that emit low energy photons \citep{Tsygankov07}.

The \astro\ spectrum of \src\ resembles the typical spectra of accretion powered pulsar. It is often described with absorbed power law continuum for the low energy part of the energy spectrum plus high energy exponential cut-off model for the high energy part. In addition, a black body component is often sued to account for soft excess in the spectrum and Gaussian components for emission lines \citep{White83}. The flux level during the 2021 outburst was similar to that during 2018 outburst \citep{Serim22, Mandal22}. But the spectral parameters show significant differences. 

The inclusion of thermal black body component in \src\ spectra is debatable and needs more detailed study during future outburst of the source. This is because, the black body component was not included in spectra of 1999 and 2009 outburst \citep{Inam04, Gupta18}. \citet{Serim22} modelled the 0.8-12.0 keV \textit{NICER} spectra of 2018 outburst with photoelectric absorption and power law with a high-energy cut-off model, and a Gaussian component to characterize the iron emission line \citep{Finger96, Inam04}. The $\sim$150 d spectrum did not require a thermal black body component. Even the time averaged spectra of about 20 d near the peak resulted in a black body temperature $\sim$0.56 keV, which was much less than $\sim$0.95 keV reported by \citep{Gupta19} with 0.9-79.0 keV \textit{NuSTAR} observations of 2018. In fact, \citet{Serim22} were unable to resolve the black body component in most of the \textit{NICER} observations. Moreover, inconsistent photon indices were obtained when black body component was included in the spectral model. 

During 2021 outburst, from 0.8-12.0 keV \textit{NICER} data, \citet{Mandal22} reported a black body temperature of $\sim$0.25 keV which is again much less than that observed in previous works. In the current work, the black body temperature was estimated to be 1.61 keV. Such a high black body temperature was once reported during quiescence with \textit{Chandra} data \citep{Tsygankov17} and it was interpreted as a result of continued accretion from the cold accretion disc. Removal of this component (Model \textit{M3}, shown in Figure~\ref{fig:spec}) did not give consistent results. The best fit had much smaller spectral index (Table~\ref{tab:spec} - Model \textit{M3}). In some Be X-ray binaries, additional absorbing component is required to bring out an adequate spectral continuum \citep{Naik11, Jaisawal16}. Therefore, instead of black body, including additional component (\texttt{pcfabs}) to characterize the absorbing material in the vicinity of the source gave an equally good spectral fit (Table~\ref{tab:spec} - Model \textit{M4}). 

The photon index ($\Gamma$) is known to decrease with the source luminosity. In the current work, $\Gamma\sim$0.28 has been observed in Model \textit{M2}, which is less than that during 2021 \textit{NICER} observation at slightly higher flux level ($\Gamma\sim$0.4). Model \textit{M4} gave a higher $\Gamma\sim$0.53. Compared to previous outbursts, where \citet{Gupta18} and \citet{Serim22} reported $\Gamma\sim$0.4 at higher flux level while \citet{Inam04} reported much higher $\Gamma$ at lower flux levels, it appears that Model \textit{M4} better describes the spectrum of \src. \citet{Mandal22} reported a transition from sub-critical to super-critical regime just after the \astro\ observation. This could be the reason for differences in the pulse profile and ambiguous spectral index and blackbody temperature. Continuous monitoring of this transient pulsar at different phases of future outbursts is therefore encouraged to bring about a complete explanation of timing and spectral variability.

Another interesting spectral component seen in \src\ is the emission feature at $\sim$0.83 keV. This line has never been reported in this source and could be an instrumental effect. Nevertheless, it is worth while to mention that occurrence of emission lines over a spectral continuum is a complex phenomena and crucially depends on contribution from X-ray irradiated stellar wind (in HMXB pulsars), accretion disk (mostly in low-mass X-ray binary sources), accretion column and column density \citep{Jimenez05, Rodes11}. High-resolution spectroscopic studies of X-ray binaries with \textit{Chandra} \citep{Canizares05} and \textit{XMM-Newton} \citep{den01} have shown presence of extended outflows (or X-ray emitting plasma) near the accretion disk. The presence of 0.83 keV emission line in case of \src\ could be a signature of unresolved iron line complex (Fe\texttt{XVII} - Fe\texttt{XIX}) observed at about 10 million Kelvin plasma temperatures \citep{Parenti17}. Several HMXBs such as LMC X-4 \citep{Negueruela98, Sharma23}, 4U 1538-52 \citep{Rodes11} and 4U 1700-37 \citep{Boroson03, Mart21} have well established soft X-ray line spectra. Since Be-type stars constitute a major fraction of HMXBs, therefore, in addition to high resolution X-ray studies, coordinated multi-wavelength study of Be stars should be performed for better understanding of complete accretion process in these system.

\section*{Acknowledgements}

This work has made use of software provided by the High Energy Astrophysics Science Archive Research Center (HEASARC), which is a service of the Astrophysics Science Division at NASA/GSFC. Data from ISRO's \astro\ mission, archived at the Indian Space Science Data Centre (ISSDC) has been used. The author is thankful to the LAXPC Payload Operation Center (POC) at TIFR, Mumbai for providing necessary software tools. This work has used the data from the Soft X-ray Telescope (SXT) developed at TIFR, Mumbai, and the SXT POC at TIFR is thanked for verifying and releasing the data via the ISSDC data archive and providing the necessary software tools. This research has also made use public data from the \textit{Swift}-BAT archive and MAXI data provided by RIKEN, JAXA and the MAXI team \citep{Matsuoka09}. The author acknowledges the assistance received from SERB–DST grant (CRG/2023/000043). Gratitude towards the anonymous referee for providing useful suggestions that have improved the content of this manuscript considerably.

\section{Data Availability}
Data used in this work can be accessed through the Indian Space Science Data Center (ISSDC) at https://astrobrowse.issdc.gov.in/astro\_archive/archive/Home.jsp
%
\balance
\bibliography{astroph.bib}{}

\begin{thebibliography}{}
\makeatletter
\relax
\def\mn@urlcharsother{\let\do\@makeother \do\$\do\&\do\#\do\^\do\_\do\%\do\~}
\def\mn@doi{\begingroup\mn@urlcharsother \@ifnextchar [ {\mn@doi@} {\mn@doi@[]}}
\def\mn@doi@[#1]#2{\def\@tempa{#1}\ifx\@tempa\@empty \href {http://dx.doi.org/#2} {doi:#2}\else \href {http://dx.doi.org/#2} {#1}\fi \endgroup}
\def\mn@eprint#1#2{\mn@eprint@#1:#2::\@nil}
\def\mn@eprint@arXiv#1{\href {http://arxiv.org/abs/#1} {{\tt arXiv:#1}}}
\def\mn@eprint@dblp#1{\href {http://dblp.uni-trier.de/rec/bibtex/#1.xml} {dblp:#1}}
\def\mn@eprint@#1:#2:#3:#4\@nil{\def\@tempa {#1}\def\@tempb {#2}\def\@tempc {#3}\ifx \@tempc \@empty \let \@tempc \@tempb \let \@tempb \@tempa \fi \ifx \@tempb \@empty \def\@tempb {arXiv}\fi \@ifundefined {mn@eprint@\@tempb}{\@tempb:\@tempc}{\expandafter \expandafter \csname mn@eprint@\@tempb\endcsname \expandafter{\@tempc}}}

\bibitem[\protect\citeauthoryear{{Agrawal}}{{Agrawal}}{2006}]{Agrawal06}
{Agrawal} P.~C.,  2006, \mn@doi [Advances in Space Research] {10.1016/j.asr.2006.03.038}, \href {https://ui.adsabs.harvard.edu/abs/2006AdSpR..38.2989A} {38, 2989}

\bibitem[\protect\citeauthoryear{{Agrawal} et~al.,}{{Agrawal} et~al.}{2017}]{Agrawal17}
{Agrawal} P.~C.,  et~al., 2017, \mn@doi [Journal of Astrophysics and Astronomy] {10.1007/s12036-017-9451-z}, \href {https://ui.adsabs.harvard.edu/abs/2017JApA...38...30A} {38, 30}

\bibitem[\protect\citeauthoryear{{Antia} et~al.,}{{Antia} et~al.}{2017}]{Antia17}
{Antia} H.~M.,  et~al., 2017, \mn@doi [\apjs] {10.3847/1538-4365/aa7a0e}, \href {https://ui.adsabs.harvard.edu/abs/2017ApJS..231...10A} {231, 10}

\bibitem[\protect\citeauthoryear{{Antia} et~al.,}{{Antia} et~al.}{2021}]{Antia21}
{Antia} H.~M.,  et~al., 2021, \mn@doi [Journal of Astrophysics and Astronomy] {10.1007/s12036-021-09712-8}, \href {https://ui.adsabs.harvard.edu/abs/2021JApA...42...32A} {42, 32}

\bibitem[\protect\citeauthoryear{{Apparao}, {Naranan}, {Kelley}  \& {Bradt}}{{Apparao} et~al.}{1980}]{Apparao80}
{Apparao} K.~M.~V.,  {Naranan} S.,  {Kelley} R.~L.,   {Bradt} H.~V.,  1980, \aap, \href {https://ui.adsabs.harvard.edu/abs/1980A&A....89..249A} {89, 249}

\bibitem[\protect\citeauthoryear{{Arnaud}}{{Arnaud}}{1996}]{Arnaud96}
{Arnaud} K.~A.,  1996, in {Jacoby} G.~H.,  {Barnes} J.,  eds,  Astronomical Society of the Pacific Conference Series Vol. 101, Astronomical Data Analysis Software and Systems V. p.~17

\bibitem[\protect\citeauthoryear{{Barthelmy} et~al.,}{{Barthelmy} et~al.}{2005}]{Barthelmy05}
{Barthelmy} S.~D.,  et~al., 2005, \mn@doi [\ssr] {10.1007/s11214-005-5096-3}, \href {https://ui.adsabs.harvard.edu/abs/2005SSRv..120..143B} {120, 143}

\bibitem[\protect\citeauthoryear{{Basko} \& {Sunyaev}}{{Basko} \& {Sunyaev}}{1976}]{Basko76}
{Basko} M.~M.,  {Sunyaev} R.~A.,  1976, \mn@doi [\mnras] {10.1093/mnras/175.2.395}, \href {https://ui.adsabs.harvard.edu/abs/1976MNRAS.175..395B} {175, 395}

\bibitem[\protect\citeauthoryear{{Baykal}, {Stark}  \& {Swank}}{{Baykal} et~al.}{2002}]{Baykal02}
{Baykal} A.,  {Stark} M.~J.,   {Swank} J.~H.,  2002, \mn@doi [\apj] {10.1086/339429}, \href {https://ui.adsabs.harvard.edu/abs/2002ApJ...569..903B} {569, 903}

\bibitem[\protect\citeauthoryear{{Becker} \& {Wolff}}{{Becker} \& {Wolff}}{2007}]{Becker07}
{Becker} P.~A.,  {Wolff} M.~T.,  2007, \mn@doi [\apj] {10.1086/509108}, \href {https://ui.adsabs.harvard.edu/abs/2007ApJ...654..435B} {654, 435}

\bibitem[\protect\citeauthoryear{{Beklen}, {Finger}  \& {GBM Pulsar Project Team}}{{Beklen} et~al.}{2009}]{Beklen09}
{Beklen} E.,  {Finger} M.~H.,   {GBM Pulsar Project Team} 2009, The Astronomer's Telegram, \href {https://ui.adsabs.harvard.edu/abs/2009ATel.2275....1B} {2275, 1}

\bibitem[\protect\citeauthoryear{{Beri}, {Sharma}, {Roy}, {Gaur}, {Altamirano}, {Andersson}, {Gittins}  \& {Celora}}{{Beri} et~al.}{2023}]{Beri23}
{Beri} A.,  {Sharma} R.,  {Roy} P.,  {Gaur} V.,  {Altamirano} D.,  {Andersson} N.,  {Gittins} F.,   {Celora} T.,  2023, \mn@doi [\mnras] {10.1093/mnras/stad902}, \href {https://ui.adsabs.harvard.edu/abs/2023MNRAS.521.5904B} {521, 5904}

\bibitem[\protect\citeauthoryear{{Bhattacharya}}{{Bhattacharya}}{2017}]{Bhattacharya17}
{Bhattacharya} D.,  2017, \mn@doi [Journal of Astrophysics and Astronomy] {10.1007/s12036-017-9461-x}, \href {https://ui.adsabs.harvard.edu/abs/2017JApA...38...51B} {38, 51}

\bibitem[\protect\citeauthoryear{{Bildsten} et~al.,}{{Bildsten} et~al.}{1997}]{Bildsten97}
{Bildsten} L.,  et~al., 1997, \mn@doi [\apjs] {10.1086/313060}, \href {https://ui.adsabs.harvard.edu/abs/1997ApJS..113..367B} {113, 367}

\bibitem[\protect\citeauthoryear{{Blackburn}, {Shaw}, {Payne}, {Hayes}  \& {Heasarc}}{{Blackburn} et~al.}{1999}]{Blackburn99}
{Blackburn} J.~K.,  {Shaw} R.~A.,  {Payne} H.~E.,  {Hayes} J.~J.~E.,   {Heasarc} 1999, {FTOOLS: A general package of software to manipulate FITS files}, Astrophysics Source Code Library, record ascl:9912.002

\bibitem[\protect\citeauthoryear{{Boldin}, {Tsygankov}  \& {Lutovinov}}{{Boldin} et~al.}{2013}]{Boldin13}
{Boldin} P.~A.,  {Tsygankov} S.~S.,   {Lutovinov} A.~A.,  2013, \mn@doi [Astronomy Letters] {10.1134/S1063773713060029}, \href {https://ui.adsabs.harvard.edu/abs/2013AstL...39..375B} {39, 375}

\bibitem[\protect\citeauthoryear{{Boroson}, {Vrtilek}, {Kallman}  \& {Corcoran}}{{Boroson} et~al.}{2003}]{Boroson03}
{Boroson} B.,  {Vrtilek} S.~D.,  {Kallman} T.,   {Corcoran} M.,  2003, \mn@doi [\apj] {10.1086/375636}, \href {https://ui.adsabs.harvard.edu/abs/2003ApJ...592..516B} {592, 516}

\bibitem[\protect\citeauthoryear{{Caballero} et~al.,}{{Caballero} et~al.}{2007}]{Caballero07}
{Caballero} I.,  et~al., 2007, \mn@doi [\aap] {10.1051/0004-6361:20067032}, \href {https://ui.adsabs.harvard.edu/abs/2007A&A...465L..21C} {465, L21}

\bibitem[\protect\citeauthoryear{{Canizares} et~al.,}{{Canizares} et~al.}{2005}]{Canizares05}
{Canizares} C.~R.,  et~al., 2005, \mn@doi [\pasp] {10.1086/432898}, \href {https://ui.adsabs.harvard.edu/abs/2005PASP..117.1144C} {117, 1144}

\bibitem[\protect\citeauthoryear{{Devasia}, {James}, {Paul}  \& {Indulekha}}{{Devasia} et~al.}{2011a}]{Devasia11}
{Devasia} J.,  {James} M.,  {Paul} B.,   {Indulekha} K.,  2011a, \mn@doi [\mnras] {10.1111/j.1365-2966.2011.18407.x}, \href {https://ui.adsabs.harvard.edu/abs/2011MNRAS.414.1023D} {414, 1023}

\bibitem[\protect\citeauthoryear{{Devasia}, {James}, {Paul}  \& {Indulekha}}{{Devasia} et~al.}{2011b}]{Devasia11b}
{Devasia} J.,  {James} M.,  {Paul} B.,   {Indulekha} K.,  2011b, \mn@doi [\mnras] {10.1111/j.1365-2966.2011.19269.x}, \href {https://ui.adsabs.harvard.edu/abs/2011MNRAS.417..348D} {417, 348}

\bibitem[\protect\citeauthoryear{{Finger}, {Wilson}  \& {Chakrabarty}}{{Finger} et~al.}{1996}]{Finger96}
{Finger} M.~H.,  {Wilson} R.~B.,   {Chakrabarty} D.,  1996, \aaps, \href {https://ui.adsabs.harvard.edu/abs/1996A&AS..120C.209F} {120, 209}

\bibitem[\protect\citeauthoryear{{Grindlay}, {Petro}  \& {McClintock}}{{Grindlay} et~al.}{1984}]{Grindlay84}
{Grindlay} J.~E.,  {Petro} L.~D.,   {McClintock} J.~E.,  1984, \mn@doi [\apj] {10.1086/161650}, \href {https://ui.adsabs.harvard.edu/abs/1984ApJ...276..621G} {276, 621}

\bibitem[\protect\citeauthoryear{{Gupta}, {Naik}, {Jaisawal}  \& {Epili}}{{Gupta} et~al.}{2018}]{Gupta18}
{Gupta} S.,  {Naik} S.,  {Jaisawal} G.~K.,   {Epili} P.~R.,  2018, \mn@doi [\mnras] {10.1093/mnras/sty1804}, \href {https://ui.adsabs.harvard.edu/abs/2018MNRAS.479.5612G} {479, 5612}

\bibitem[\protect\citeauthoryear{{Gupta}, {Naik}  \& {Jaisawal}}{{Gupta} et~al.}{2019}]{Gupta19}
{Gupta} S.,  {Naik} S.,   {Jaisawal} G.~K.,  2019, \mn@doi [\mnras] {10.1093/mnras/stz2795}, \href {https://ui.adsabs.harvard.edu/abs/2019MNRAS.490.2458G} {490, 2458}

\bibitem[\protect\citeauthoryear{{HI4PI Collaboration} et~al.,}{{HI4PI Collaboration} et~al.}{2016}]{nh16}
{HI4PI Collaboration} et~al., 2016, \mn@doi [\aap] {10.1051/0004-6361/201629178}, \href {https://ui.adsabs.harvard.edu/abs/2016A&A...594A.116H} {594, A116}

\bibitem[\protect\citeauthoryear{{Hazra}, {Pal}, {Mandal}  \& {Bhunia}}{{Hazra} et~al.}{2021}]{Hazra21}
{Hazra} M.,  {Pal} S.,  {Mandal} M.,   {Bhunia} B.,  2021, The Astronomer's Telegram, \href {https://ui.adsabs.harvard.edu/abs/2021ATel14349....1H} {14349, 1}

\bibitem[\protect\citeauthoryear{{{\.I}nam}, {Baykal}, {Matthew Scott}, {Finger}  \& {Swank}}{{{\.I}nam} et~al.}{2004}]{Inam04}
{{\.I}nam} S.~{\c{C}}.,  {Baykal} A.,  {Matthew Scott} D.,  {Finger} M.,   {Swank} J.,  2004, \mn@doi [\mnras] {10.1111/j.1365-2966.2004.07478.x}, \href {https://ui.adsabs.harvard.edu/abs/2004MNRAS.349..173I} {349, 173}

\bibitem[\protect\citeauthoryear{{Jain}, {Paul}  \& {Dutta}}{{Jain} et~al.}{2009}]{Jain09}
{Jain} C.,  {Paul} B.,   {Dutta} A.,  2009, \mn@doi [Research in Astronomy and Astrophysics] {10.1088/1674-4527/9/12/002}, \href {https://ui.adsabs.harvard.edu/abs/2009RAA.....9.1303J} {9, 1303}

\bibitem[\protect\citeauthoryear{{Jaisawal}, {Naik}  \& {Epili}}{{Jaisawal} et~al.}{2016}]{Jaisawal16}
{Jaisawal} G.~K.,  {Naik} S.,   {Epili} P.,  2016, \mn@doi [\mnras] {10.1093/mnras/stw085}, \href {https://ui.adsabs.harvard.edu/abs/2016MNRAS.457.2749J} {457, 2749}

\bibitem[\protect\citeauthoryear{{Jaisawal}, {Naik}  \& {Chenevez}}{{Jaisawal} et~al.}{2018}]{Jaisawal18}
{Jaisawal} G.~K.,  {Naik} S.,   {Chenevez} J.,  2018, \mn@doi [\mnras] {10.1093/mnras/stx3082}, \href {https://ui.adsabs.harvard.edu/abs/2018MNRAS.474.4432J} {474, 4432}

\bibitem[\protect\citeauthoryear{{Ji} et~al.,}{{Ji} et~al.}{2020}]{Ji20}
{Ji} L.,  et~al., 2020, \mn@doi [\mnras] {10.1093/mnras/stz2745}, \href {https://ui.adsabs.harvard.edu/abs/2020MNRAS.491.1851J} {491, 1851}

\bibitem[\protect\citeauthoryear{{Jimenez-Garate}, {Raymond}, {Liedahl}  \& {Hailey}}{{Jimenez-Garate} et~al.}{2005}]{Jimenez05}
{Jimenez-Garate} M.~A.,  {Raymond} J.~C.,  {Liedahl} D.~A.,   {Hailey} C.~J.,  2005, \mn@doi [\apj] {10.1086/426702}, \href {https://ui.adsabs.harvard.edu/abs/2005ApJ...625..931J} {625, 931}

\bibitem[\protect\citeauthoryear{{Kelley}, {Apparao}, {Doxsey}, {Jernigan}, {Naranan}  \& {Rappaport}}{{Kelley} et~al.}{1981}]{Kelley81}
{Kelley} R.~L.,  {Apparao} K.~M.~V.,  {Doxsey} R.~E.,  {Jernigan} J.~G.,  {Naranan} S.,   {Rappaport} S.,  1981, \mn@doi [\apj] {10.1086/158591}, \href {https://ui.adsabs.harvard.edu/abs/1981ApJ...243..251K} {243, 251}

\bibitem[\protect\citeauthoryear{{Klochkov}, {Santangelo}, {Staubert}  \& {Ferrigno}}{{Klochkov} et~al.}{2008}]{Klochkov08}
{Klochkov} D.,  {Santangelo} A.,  {Staubert} R.,   {Ferrigno} C.,  2008, \mn@doi [\aap] {10.1051/0004-6361:200810673}, \href {https://ui.adsabs.harvard.edu/abs/2008A&A...491..833K} {491, 833}

\bibitem[\protect\citeauthoryear{{Krimm} et~al.,}{{Krimm} et~al.}{2013}]{Krimm13}
{Krimm} H.~A.,  et~al., 2013, \mn@doi [\apjs] {10.1088/0067-0049/209/1/14}, \href {https://ui.adsabs.harvard.edu/abs/2013ApJS..209...14K} {209, 14}

\bibitem[\protect\citeauthoryear{{Leahy}}{{Leahy}}{1987}]{Leahy87}
{Leahy} D.~A.,  1987, \aap, \href {https://ui.adsabs.harvard.edu/abs/1987A&A...180..275L} {180, 275}

\bibitem[\protect\citeauthoryear{{Leahy}, {Darbro}, {Elsner}, {Weisskopf}, {Sutherland}, {Kahn}  \& {Grindlay}}{{Leahy} et~al.}{1983}]{Leahy83}
{Leahy} D.~A.,  {Darbro} W.,  {Elsner} R.~F.,  {Weisskopf} M.~C.,  {Sutherland} P.~G.,  {Kahn} S.,   {Grindlay} J.~E.,  1983, \mn@doi [\apj] {10.1086/160766}, \href {https://ui.adsabs.harvard.edu/abs/1983ApJ...266..160L} {266, 160}

\bibitem[\protect\citeauthoryear{{Lutovinov} \& {Tsygankov}}{{Lutovinov} \& {Tsygankov}}{2009}]{Lutovinov09}
{Lutovinov} A.~A.,  {Tsygankov} S.~S.,  2009, \mn@doi [Astronomy Letters] {10.1134/S1063773709070019}, \href {https://ui.adsabs.harvard.edu/abs/2009AstL...35..433L} {35, 433}

\bibitem[\protect\citeauthoryear{{Mandal} \& {Pal}}{{Mandal} \& {Pal}}{2022}]{Mandal22}
{Mandal} M.,  {Pal} S.,  2022, \mn@doi [\apss] {10.1007/s10509-022-04150-6}, \href {https://ui.adsabs.harvard.edu/abs/2022Ap&SS.367..112M} {367, 112}

\bibitem[\protect\citeauthoryear{{Mart{\'\i}nez-Chicharro} et~al.,}{{Mart{\'\i}nez-Chicharro} et~al.}{2021}]{Mart21}
{Mart{\'\i}nez-Chicharro} M.,  et~al., 2021, \mn@doi [\mnras] {10.1093/mnras/staa3956}, \href {https://ui.adsabs.harvard.edu/abs/2021MNRAS.501.5646M} {501, 5646}

\bibitem[\protect\citeauthoryear{{Matsuoka} et~al.,}{{Matsuoka} et~al.}{2009}]{Matsuoka09}
{Matsuoka} M.,  et~al., 2009, \mn@doi [\pasj] {10.1093/pasj/61.5.999}, \href {https://ui.adsabs.harvard.edu/abs/2009PASJ...61..999M} {61, 999}

\bibitem[\protect\citeauthoryear{{Mihara} et~al.,}{{Mihara} et~al.}{2011}]{Mihara11}
{Mihara} T.,  et~al., 2011, \mn@doi [\pasj] {10.1093/pasj/63.sp3.S623}, \href {https://ui.adsabs.harvard.edu/abs/2011PASJ...63S.623M} {63, S623}

\bibitem[\protect\citeauthoryear{{Mukherjee} \& {Paul}}{{Mukherjee} \& {Paul}}{2005}]{Mukherjee5}
{Mukherjee} U.,  {Paul} B.,  2005, \mn@doi [\aap] {10.1051/0004-6361:20041665}, \href {https://ui.adsabs.harvard.edu/abs/2005A&A...431..667M} {431, 667}

\bibitem[\protect\citeauthoryear{{Naik}, {Paul}, {Kachhara}  \& {Vadawale}}{{Naik} et~al.}{2011}]{Naik11}
{Naik} S.,  {Paul} B.,  {Kachhara} C.,   {Vadawale} S.~V.,  2011, \mn@doi [\mnras] {10.1111/j.1365-2966.2010.18128.x}, \href {https://ui.adsabs.harvard.edu/abs/2011MNRAS.413..241N} {413, 241}

\bibitem[\protect\citeauthoryear{{Negueruela}}{{Negueruela}}{1998}]{Negueruela98}
{Negueruela} I.,  1998, \mn@doi [\aap] {10.48550/arXiv.astro-ph/9807158}, \href {https://ui.adsabs.harvard.edu/abs/1998A&A...338..505N} {338, 505}

\bibitem[\protect\citeauthoryear{{Parenti}, {del Zanna}, {Petralia}, {Reale}, {Teriaca}, {Testa}  \& {Mason}}{{Parenti} et~al.}{2017}]{Parenti17}
{Parenti} S.,  {del Zanna} G.,  {Petralia} A.,  {Reale} F.,  {Teriaca} L.,  {Testa} P.,   {Mason} H.~E.,  2017, \mn@doi [\apj] {10.3847/1538-4357/aa835f}, \href {https://ui.adsabs.harvard.edu/abs/2017ApJ...846...25P} {846, 25}

\bibitem[\protect\citeauthoryear{{Parmar}, {White}, {Stella}, {Izzo}  \& {Ferri}}{{Parmar} et~al.}{1989}]{Parmar89}
{Parmar} A.~N.,  {White} N.~E.,  {Stella} L.,  {Izzo} C.,   {Ferri} P.,  1989, \mn@doi [\apj] {10.1086/167204}, \href {https://ui.adsabs.harvard.edu/abs/1989ApJ...338..359P} {338, 359}

\bibitem[\protect\citeauthoryear{{Porter} \& {Rivinius}}{{Porter} \& {Rivinius}}{2003}]{Porter03}
{Porter} J.~M.,  {Rivinius} T.,  2003, \mn@doi [\pasp] {10.1086/378307}, \href {https://ui.adsabs.harvard.edu/abs/2003PASP..115.1153P} {115, 1153}

\bibitem[\protect\citeauthoryear{{Raichur} \& {Paul}}{{Raichur} \& {Paul}}{2010}]{Raichur10}
{Raichur} H.,  {Paul} B.,  2010, \mn@doi [\mnras] {10.1111/j.1365-2966.2010.16862.x}, \href {https://ui.adsabs.harvard.edu/abs/2010MNRAS.406.2663R} {406, 2663}

\bibitem[\protect\citeauthoryear{{Ramadevi} et~al.,}{{Ramadevi} et~al.}{2018}]{Ramadevi18}
{Ramadevi} M.~C.,  et~al., 2018, \mn@doi [Journal of Astrophysics and Astronomy] {10.1007/s12036-017-9506-1}, \href {https://ui.adsabs.harvard.edu/abs/2018JApA...39...11R} {39, 11}

\bibitem[\protect\citeauthoryear{{Rao}, {Bhattacharya}, {Bhalerao}, {Vadawale}  \& {Sreekumar}}{{Rao} et~al.}{2017}]{Rao17}
{Rao} A.~R.,  {Bhattacharya} D.,  {Bhalerao} V.~B.,  {Vadawale} S.~V.,   {Sreekumar} S.,  2017, \mn@doi [Current Science] {10.18520/cs/v113/i04/595-598}, \href {https://ui.adsabs.harvard.edu/abs/2017CSci..113..595R} {113, 595}

\bibitem[\protect\citeauthoryear{{Reig}}{{Reig}}{2011}]{Reig11}
{Reig} P.,  2011, \mn@doi [\apss] {10.1007/s10509-010-0575-8}, \href {https://ui.adsabs.harvard.edu/abs/2011Ap&SS.332....1R} {332, 1}

\bibitem[\protect\citeauthoryear{{Rivinius}, {Carciofi}  \& {Martayan}}{{Rivinius} et~al.}{2013}]{Rivinius13}
{Rivinius} T.,  {Carciofi} A.~C.,   {Martayan} C.,  2013, \mn@doi [\aapr] {10.1007/s00159-013-0069-0}, \href {https://ui.adsabs.harvard.edu/abs/2013A&ARv..21...69R} {21, 69}

\bibitem[\protect\citeauthoryear{{Rodes-Roca}, {Page}, {Torrej{\'o}n}, {Osborne}  \& {Bernab{\'e}u}}{{Rodes-Roca} et~al.}{2011}]{Rodes11}
{Rodes-Roca} J.~J.,  {Page} K.~L.,  {Torrej{\'o}n} J.~M.,  {Osborne} J.~P.,   {Bernab{\'e}u} G.,  2011, \mn@doi [\aap] {10.1051/0004-6361/201014324}, \href {https://ui.adsabs.harvard.edu/abs/2011A&A...526A..64R} {526, A64}

\bibitem[\protect\citeauthoryear{{Sakamoto} et~al.,}{{Sakamoto} et~al.}{2016}]{Sakamoto16}
{Sakamoto} T.,  et~al., 2016, \mn@doi [\pasj] {10.1093/pasj/psv130}, \href {https://ui.adsabs.harvard.edu/abs/2016PASJ...68S...2S} {68, S2}

\bibitem[\protect\citeauthoryear{{Serim}, {{\"O}z{\"u}do{\u{g}}ru}, {D{\"o}nmez}, {{\c{S}}ahiner}, {Serim}, {Baykal}  \& {{\.I}nam}}{{Serim} et~al.}{2022}]{Serim22}
{Serim} M.~M.,  {{\"O}z{\"u}do{\u{g}}ru} {\"O}.~C.,  {D{\"o}nmez} {\c{C}}.~K.,  {{\c{S}}ahiner} {\c{S}}.,  {Serim} D.,  {Baykal} A.,   {{\.I}nam} S.~{\c{C}}.,  2022, \mn@doi [\mnras] {10.1093/mnras/stab3547}, \href {https://ui.adsabs.harvard.edu/abs/2022MNRAS.510.1438S} {510, 1438}

\bibitem[\protect\citeauthoryear{{Sharma}, {Jain}, {Rikame}  \& {Paul}}{{Sharma} et~al.}{2023}]{Sharma23}
{Sharma} R.,  {Jain} C.,  {Rikame} K.,   {Paul} B.,  2023, \mn@doi [\mnras] {10.1093/mnras/stac3572}, \href {https://ui.adsabs.harvard.edu/abs/2023MNRAS.519.1764S} {519, 1764}

\bibitem[\protect\citeauthoryear{{Shrader}, {Sutaria}, {Singh}  \& {Macomb}}{{Shrader} et~al.}{1999}]{Shrader99}
{Shrader} C.~R.,  {Sutaria} F.~K.,  {Singh} K.~P.,   {Macomb} D.~J.,  1999, \mn@doi [\apj] {10.1086/306785}, \href {https://ui.adsabs.harvard.edu/abs/1999ApJ...512..920S} {512, 920}

\bibitem[\protect\citeauthoryear{{Singh} et~al.,}{{Singh} et~al.}{2014}]{Singh14}
{Singh} K.~P.,  et~al., 2014, in {Takahashi} T.,  {den Herder} J.-W.~A.,   {Bautz} M.,  eds,  Society of Photo-Optical Instrumentation Engineers (SPIE) Conference Series Vol. 9144, Space Telescopes and Instrumentation 2014: Ultraviolet to Gamma Ray. p. 91441S, \mn@doi{10.1117/12.2062667}

\bibitem[\protect\citeauthoryear{{Singh} et~al.,}{{Singh} et~al.}{2016}]{Singh16}
{Singh} K.~P.,  et~al., 2016, in {den Herder} J.-W.~A.,  {Takahashi} T.,   {Bautz} M.,  eds,  Society of Photo-Optical Instrumentation Engineers (SPIE) Conference Series Vol. 9905, Space Telescopes and Instrumentation 2016: Ultraviolet to Gamma Ray. p. 99051E, \mn@doi{10.1117/12.2235309}

\bibitem[\protect\citeauthoryear{{Singh} et~al.,}{{Singh} et~al.}{2017}]{Singh17}
{Singh} K.~P.,  et~al., 2017, \mn@doi [Journal of Astrophysics and Astronomy] {10.1007/s12036-017-9448-7}, \href {https://ui.adsabs.harvard.edu/abs/2017JApA...38...29S} {38, 29}

\bibitem[\protect\citeauthoryear{{Sturner} \& {Dermer}}{{Sturner} \& {Dermer}}{1994}]{Sturner94}
{Sturner} S.~J.,  {Dermer} C.~D.,  1994, \aap, \href {https://ui.adsabs.harvard.edu/abs/1994A&A...284..161S} {284, 161}

\bibitem[\protect\citeauthoryear{{Tandon} et~al.,}{{Tandon} et~al.}{2017}]{Tandon17}
{Tandon} S.~N.,  et~al., 2017, \mn@doi [\aj] {10.3847/1538-3881/aa8451}, \href {https://ui.adsabs.harvard.edu/abs/2017AJ....154..128T} {154, 128}

\bibitem[\protect\citeauthoryear{{Tsygankov}, {Lutovinov}, {Churazov}  \& {Sunyaev}}{{Tsygankov} et~al.}{2007}]{Tsygankov07}
{Tsygankov} S.~S.,  {Lutovinov} A.~A.,  {Churazov} E.~M.,   {Sunyaev} R.~A.,  2007, \mn@doi [Astronomy Letters] {10.1134/S1063773707060023}, \href {https://ui.adsabs.harvard.edu/abs/2007AstL...33..368T} {33, 368}

\bibitem[\protect\citeauthoryear{{Tsygankov}, {Wijnands}, {Lutovinov}, {Degenaar}  \& {Poutanen}}{{Tsygankov} et~al.}{2017}]{Tsygankov17}
{Tsygankov} S.~S.,  {Wijnands} R.,  {Lutovinov} A.~A.,  {Degenaar} N.,   {Poutanen} J.,  2017, \mn@doi [\mnras] {10.1093/mnras/stx1255}, \href {https://ui.adsabs.harvard.edu/abs/2017MNRAS.470..126T} {470, 126}

\bibitem[\protect\citeauthoryear{{Verner}, {Ferland}, {Korista}  \& {Yakovlev}}{{Verner} et~al.}{1996}]{Verner96}
{Verner} D.~A.,  {Ferland} G.~J.,  {Korista} K.~T.,   {Yakovlev} D.~G.,  1996, \mn@doi [\apj] {10.1086/177435}, \href {https://ui.adsabs.harvard.edu/abs/1996ApJ...465..487V} {465, 487}

\bibitem[\protect\citeauthoryear{{White}, {Swank}  \& {Holt}}{{White} et~al.}{1983}]{White83}
{White} N.~E.,  {Swank} J.~H.,   {Holt} S.~S.,  1983, \mn@doi [\apj] {10.1086/161162}, \href {https://ui.adsabs.harvard.edu/abs/1983ApJ...270..711W} {270, 711}

\bibitem[\protect\citeauthoryear{{Wilms}, {Allen}  \& {McCray}}{{Wilms} et~al.}{2000}]{Wilms00}
{Wilms} J.,  {Allen} A.,   {McCray} R.,  2000, \mn@doi [\apj] {10.1086/317016}, \href {https://ui.adsabs.harvard.edu/abs/2000ApJ...542..914W} {542, 914}

\bibitem[\protect\citeauthoryear{{Yadav} et~al.,}{{Yadav} et~al.}{2016}]{Yadav16}
{Yadav} J.~S.,  et~al., 2016, in {den Herder} J.-W.~A.,  {Takahashi} T.,   {Bautz} M.,  eds,  Society of Photo-Optical Instrumentation Engineers (SPIE) Conference Series Vol. 9905, Space Telescopes and Instrumentation 2016: Ultraviolet to Gamma Ray. p. 99051D, \mn@doi{10.1117/12.2231857}

\bibitem[\protect\citeauthoryear{{den Herder} et~al.,}{{den Herder} et~al.}{2001}]{den01}
{den Herder} J.~W.,  et~al., 2001, \mn@doi [\aap] {10.1051/0004-6361:20000058}, \href {https://ui.adsabs.harvard.edu/abs/2001A&A...365L...7D} {365, L7}

\bibitem[\protect\citeauthoryear{{in't Zand}, {Corbet}  \& {Marshall}}{{in't Zand} et~al.}{2001}]{Zand01}
{in't Zand} J.~J.~M.,  {Corbet} R.~H.~D.,   {Marshall} F.~E.,  2001, \mn@doi [\apjl] {10.1086/320688}, \href {https://ui.adsabs.harvard.edu/abs/2001ApJ...553L.165I} {553, L165}

\makeatother
\end{thebibliography}
\bibliographystyle{mnras}
\end{document}